\documentclass[10pt,a4paper]{article}
\usepackage{feynmp} 
\usepackage{amsmath,amsfonts,amssymb,amsthm}

\usepackage{amsmath}

\usepackage{amssymb}
\usepackage{color}
\usepackage{colordvi}
\usepackage{amsmath}

\newtheorem{thm}{Theorem}
\newtheorem{defn}{Definition}
\newtheorem{cor}{Corollary}
\newtheorem{lem}{Lemma}

\newtheorem{prop}{Proposition}
\newtheorem{ex}{Example}
\newtheorem{rk}{Remark}

\usepackage{amssymb}

\def\dbar{d{\hskip-1pt\bar{}}\hskip1pt}
\def\cutoffint{-\hskip -10pt\int}
\def\altcutoffint{=\hskip -11pt\int}
\def \C{{\! \rm \
I \!\!\!C}}

\def \R {{\! \rm \ I \!R}}

\def \N {{\! \rm \ I \!N}}

\def \Z {{\! \rm Z\! \!Z}}
 \def\cutoffint{-\hskip -10pt\int}

\def\otherterm#1{{\it#1}}
\def \e {{\epsilon}}

\def \Ci {{C^\infty}}

\def \endsquare{ $\sqcup \!\!\!\! \sqcap$ }

\def\Ci{C^\infty}

\def\dd{\partial}

\def\Di{D\kern -.65em /}
\def\Dii{D\kern -.45em /}
\def\di{{\dd}\kern -.55em /}
\def\dii{{\dd}\kern -.40em /}

\def\noi{\noindent}

\def\to{\rightarrow}

\def\Dd{{\mathcal D}}

\def\Nn{{\mathcal N}}

\def\Rr{{\mathcal R}}

\def\={\cong}
\def\>{\supset}
\def\<{\subset}

\def\12{\frac{1}{2}}
\def\2{\Dd}
\def\3{\Nn}
\def\4{\Rr}
\def\6{\cup}
\def\8{\otimes}
\def\0{^{\circ}}
\def\){\hfill{\ \qed}\enddemo}

\def\R{\mathbb{R}}
\def\C{\mathbb{C}}

\def\e{\varepsilon}

\def\N{\NN}

\def\Si{\Sigma}

\def\Z{\ZZ}

\def\Dd{{\mathcal D}}

\def\Nn{{\mathcal N}}

\def\Rr{{\mathcal R}}

\def\Si{S\kern -.65em /}

\def \C{{\! \rm \ I \!\!\!C}}
\def \R {{\! \rm \ I \!R}}
\def \N {{\! \rm \ I \!N}}

\def \Z {{\! \rm Z\! \!Z}}
\def\cutoffint{-\hskip -12pt\int}

\def\otherterm#1{{\it#1}}
\def \e {{\epsilon}}

\def \Ci {{C^\infty}}

\begin{document}
\title{\bf Renormalised iterated integrals  of symbols with linear constraints}
\author{ Sylvie PAYCHA}
\maketitle
\section*{Abstract} Given a holomorphic regularisation procedure
(e.g. Riesz or dimensional regularisation) on
classical symbols, we define  {\it renormalised multiple  integrals} of radial
classical 
symbols  with linear constraints. To do so,  we first prove the existence  of 
meromorphic extensions of multiple integrals of holomorphic perturbations of
radial symbols with linear constraints and then implement either generalised
evaluators or a Birkhoff factorisation. Renormalised multiple integrals are
covariant and 
factorise over independent sets of constraints.
\section*{Acknowledgements}
I am very indebted to Daniel Bennequin for his comments and precious advice
while I was writing  this paper. I also very much appreciate the numerous discussions I had with
Dominique Manchon around
renormalisation  which served as a motivation to write up  this article as
well as the comments he made on a preliminary version of this paper.  Let
me further address my  
thanks  to Alessandra Frabetti for her enlightening comments on 
preliminary  drafts  of this paper, which  was completed during a stay at the
Max Planck Institute in Bonn and I also thank   Matilde Marcolli for stimulating discussions
 on parts of 
paper. I  furthermore very much benefitted from enriching discussions with Matthias
Lesch on regularisation methods for integrals of symbols  and with Mich\`ele
Vergne on  renormalisation methods  for discrete sums of symbols on integer
points of cones which helped me clarify related renormalisation issues for
multiple integrals with linear constraints. Last but not least, I would like
to thank the referee for his/her valuable comments.
\vfill \eject \noindent
\section*{Introduction}
Regularisation methods are sufficient to handle ordinary integrals arising from
one loop Feynman diagrams whereas renormalisation methods are required to
handle mutiple  integrals arising from
multiloop Feynman diagrams. Interesting algebraic constructions   have been developped to
disentangle the procedure used by physicists when computing such
integrals \cite{CK}, \cite{Kr}. Although they clarify the algebraic structure underlying the forest
formula, these algebraic approaches based on the  Hopf algebra structures on
Feynman diagrams  do not make explicit the corresponding manipulations  on the
multiple integrals. This paper aims at presenting  analytic mechanisms
underlying  renormalisation procedures  in physics on firm mathematical ground using
the language 
pseudodifferential symbols in which locality in physics translates into a factorization
property of integrals.\\ \\ We consider
 integrals of  symbols 
with  linear constraints\footnote{In the language of Feynman diagrams,
 we only deal with internal
momenta namely we integrate on all the variables.  }  that reflect the conservation of momenta; 
  properties of symbols clearly play a crucial role in 
the renormalisation procedure. When they converge we can write   such
integrals as follows:
$$\int_{\R^{nL}}\, \left(\tilde \sigma\circ B\right)(\xi_1, \cdots, \xi_L)\, d\xi_1\cdots d\xi_L, \quad \quad   
  $$  with 
    $\tilde \sigma:=  \sigma_1\otimes \cdots \otimes \sigma_I $  where
  the $\sigma_i$ are classical symbols on $\R^n$ and  $B$  an $I\times L$
matrix  of rank $L$. In the language of perturbative quantum field
theory, $n$ stands for the dimension of space time so that $n=4$, $L$ stands for the number of loops,
 $\left(\eta_1, \cdots, \eta_I\right):=B\left(\xi_1, \cdots,\xi_L\right)$ for the
internal  vertices  and the matrix  $B$ encodes  the
linear constraints they are submitted to as a result of the conservation of
momenta. To illustrate this by an example   take $I=3, L=2$,
 the symbols   $\sigma_i, i=1, 2, 3$ equal to $  \sigma(\xi)= 
\frac{1}{\left(m^2+\vert \xi\vert^2\right)^2}$ for some $m\in \R^*$ (which
introduces a mass term) and the  matrix 
$B= \left( \begin{array}{cc} 1&0\\
                                    0&1\\
				    1&1\end{array}\right).$

				    The corresponding integral   for $n=4$  reads \begin{eqnarray*}
&{}&\int_{\R^{4}}\,\int_{\R^{4}}\,\left(\left( \sigma_1\otimes \sigma_2\otimes
  \sigma_3 \right)\circ B\right)(\xi_1, \xi_2)\, 
d\xi_1\, d\xi_2\\
&{}&= \int_{\R^{4}}\,\int_{\R^{4}}\, \frac{1}{\left(m^2+\vert \xi_1\vert^2\right)^2} \, 
\frac{1}{\left(m^2+\vert \xi_2\vert^2\right)^2}\,  \frac{1}{\left(m^2+\vert \xi_1+ \xi_2\vert^2\right)^2}\, d\xi_1
\, d\xi_2.\end{eqnarray*}  \\ We wish to renormalise 
multiple 
integrals with linear constraints of this type when the integrand does not anymore lie in $L^1$    in such a way that
\begin{enumerate}
\item  the renormalised integrals coincide with
the usual integrals whenever the integrand lies in  $L^1$,
\item they satisfy a Fubini type property, i.e. are invariant under
  permutations of the variables, 
\item they factorise on disjoint sets of constraints, i.e. on products $ \left(\sigma\circ B\right)\bullet
  \left(\sigma^\prime\circ B^\prime\right):= \left(\sigma\otimes
    \sigma^\prime\right)\, \circ\,  \left(B\oplus B^\prime\right)$ where
  $\oplus$ stands for the Whitney sum. 
\end{enumerate} 
This last requirement, which would correspond in quantum field
  theory to the concatenation of Feynman diagrams, follows from the
  fundamental locality principle in physics.\\
Inspired by physicist's computations of Feynman integrals, we present two
renormalisation procedures,  a first one which uses generalised evaluators and  an alternative
method    using a Birkhoff factorisation procedure,
both of which heavily rely on meromorphicity results and both of which lead to covariant expressions.\\ \\
  Let us  briefly describe the structure of the paper. \\
Regularisation procedures for  simple integrals of symbols are  by  now  well
known and provide a precise mathematical description for  what physicists
refer to as dimensional regularisation for one loop diagrams (see
e.g. \cite{C} from a physicist's point of view and  \cite{P1} from a
mathematician's point of view  for a review of some regularisation methods used in
physics). \\ Regularisation techniques for simple integrals are  reviewed in
the first part of the paper. We describe  in   dimensional
regularisation  as an instance of more general holomorphic
regularisations and compare it with cut-off regularisation in Theorem \ref{thm:cutoffdimreg}.
 Covariance,
integration by parts and translation invariance properties are discussed in
detail in  section 3. 
 In section 4, inspired by work by Lesch and Pflaum \cite{LP} on strongly parametric
 symbols \footnote{Although our setup is different from that of strongly
   parametric symbols, it turns out that the approach of  \cite{LP} can be
 partially  adapted to our context.}, we investigate parameter dependent
integrals of symbols of the type that typically arises in the presence of external momenta in
quantum field theory. The parameter dependence in the external parameters being affine in the context of
Feynman diagrams,  we  study  
 regularised integrals  $\altcutoffint_{\R^n}\, \left(\sigma_1\otimes \cdots \otimes
 \sigma_k\right)(\xi+\eta_1, \cdots, \xi+\eta_k)\, d\xi$ which we actually
define ``modulo polynomials in the
components of the  external parameters'' $\eta_1, \cdots, \eta_k$.
Since cut-off regularised
integrals vanish on polynomials (see Proposition \ref{prop:polynomials}),  the ambiguity arising from defining the
 integrals ``up to polynomials'' is not
seen after implementing cut-off (or dimensional)  regularisation in the remaining
parameters (see Theorem  \ref{thm:parameterdeptintegrals} and  Corollary  \ref{cor:parameterdeptintegrals}).  \\
\\ The second part of the paper (sections 5-8) is dedicated to
renormalisation techniques for multiple integrals with constraints. 
In section 5 we  first renormalise multiple integrals without constraints (see
Theorem \ref{thm:holcutofftensor}) in
the spirit of previous work with D. Manchon \cite{MP}. 
The corner stone of renormalisation in our context is a meromorphicity
result, which  if fairly straightforward in the absence of constraints,
becomes  non trivial in the presence of linear constraints
\footnote{This issue is of course strongly related to the meromorphicity of Feynman integrals using dimensional regularisation
 previously  investigated by various authors from different view points see
 e.g. \cite{Sp},  \cite{CaM},\cite{E}, \cite{CM}, \cite{BW1},\cite{BW2}.}. We show (see
Theorem \ref{thm:meromultintsymb}) that the  map:
$$(z_1, \cdots, z_I) \mapsto \int_{\R^{nL}}\, \left(\tilde \sigma(\underline z)\circ B\right)(\xi_1, \cdots, \xi_L)\, d\xi_1\cdots d\xi_L, \quad \quad   
  $$  with 
    $\tilde \sigma(\underline z ):=  \sigma_1(z_1)\otimes \cdots \otimes
    \sigma_I(z_I) $ obtained from a holormorphic perturbation
  ${\cal R}: \sigma_i\mapsto \sigma_i(z)$  of {\it radial} symbols $\sigma_i$  (which can e.g. arise from dimensional
  regularisation), has a meromorphic extension 
$$\underline z \mapsto \cutoffint_{\R^{nL}}\, \left(\tilde \sigma(\underline z)\circ B\right)(\xi_1, \cdots, \xi_L)\, d\xi_1\cdots d\xi_L,  
  $$ and
  we describe its pole structure. These meromorphicity results  which are coherent with
  known results in the case of Feynman diagrams \cite{Sp}, are to our
  knowledge  new in such generality since they hold for any radial classical symbols and
  any holomorphic regularisation.\\ 
Since these meromorphic extensions coincide with ordinary multiple integrals
on the domain of holomorphicity, by analytic continuation they factorise over disjoint sets of
constraints   i.e:
$$ \cutoffint_{\R^{nL}}\, \left(\tilde \sigma(\underline
  z)\otimes \tilde \sigma^\prime(\underline
  z^\prime)\right)\circ\left( B\oplus B^\prime\right)=\left( \cutoffint_{\R^{nL}}\, \tilde
  \sigma(\underline z)\circ B\right)\, \left( \cutoffint_{\R^{nL}}\, \tilde
  \sigma^\prime(\underline z^\prime)\circ B^\prime\right),$$
whre we have set $\underline z:= (z_1, \cdots,z_I)$ and $\underline z^\prime:=
(z_1^\prime, \cdots, z_{I^\prime}^\prime)$, $\tilde \sigma:= \sigma_1\otimes
\cdots\otimes\sigma_I$, $\tilde \sigma^\prime:= \sigma^\prime_1\otimes
\cdots\otimes\sigma^\prime_{I^\prime}$, $B$ and $B^\prime$ being matrices of size
$I\times L$ and $I^\prime \times L^\prime$ respectively. \\
 We then
  describe two ways of extracting renormalised finite parts $\cutoffint_{\R^{nL}}^{{\cal R}, {\rm ren}}\tilde \sigma\circ B$ as $\underline z\to 0$ while
  preserving this factorisation property:
\begin{enumerate}
\item Using generalised evaluators (see Theorem
  \ref{thm:evaluatorsrenormalised}),
 \item Using Birkhoff factorisation (see Theorem
  \ref{thm:Birkhoffrenormalised}) after having identified
  $z_i=z$\footnote{Such an identification is natural in the context of
    dimensional regularisation by which the  dimension $n$ of the space is replaced by
    $n-z$.} and set
 the  $\sigma_i$'s to be a fixed radial symbol $\sigma$. 
\end{enumerate}
 Just as in  Connes and Kreimer's pioneering work \cite{CK}, 
in this second approach the factorisation requirement translates to a 
 character property  on a certain Hopf algebra,  the coproduct of which  reflects
the fact that one should in principle be able to perform  iterated
integrations ``packetwise'',   first integrating  on any subset of
variables and then on the remaining ones  (see \cite{BM} for  comments on this
point).   \\
As well as being multiplicative (see (\ref{eq:renormalisedfactorisation})):
$$ \cutoffint_{\R^{n(L+L^\prime)}}^{{\cal R}, {\rm ren}}\, \left(\tilde \sigma\otimes \tilde \sigma^\prime\right)\circ\left( B\oplus B^\prime\right)=\left(
\cutoffint_{\R^{nL}}^{{\cal R}, {\rm ren}}\, \tilde
  \sigma\circ B\right)\, \left( \cutoffint_{\R^{nL^\prime}}^{{\cal R}, {\rm ren}}\, \tilde
  \sigma^\prime\circ B^\prime\right),$$
renormalised multiple  integrals with constraints turn out to be covariant
(see Theorem \ref{thm:renormalisedcovariance}):
   $$\cutoffint_{\R^{nL}}^{{\cal R}, {\rm ren}}\left(\tilde \sigma\circ B\right)\circ C= \vert
   {\rm det} C\vert^{-1}\cutoffint_{\R^{nL}}^{{\cal R}, {\rm ren}}\tilde \sigma\circ B\quad \quad\forall
   C\in GL_L(\R^n)$$
   and therefore obey a Fubini property (see (\ref{eq:renormalisedFubini})):
   $$\cutoffint_{\R^{nL}}^{{\cal R}, {\rm ren}}\ \sigma\circ 
  B\left(\xi_{\rho(1)}, \cdots,
   \xi_{\rho(L)}\right)\, d\, \xi_1\cdots d\, \xi_L
   = \cutoffint_{\R^{nL}}^{{\cal R}, {\rm ren}}\ \, \sigma\circ B
   \left(\xi_{1}, \cdots,
   \xi_{L}\right)d\, \xi_1\cdots d\, \xi_L\quad \quad \forall \rho \in \Sigma_L. $$
The above factorisation property (which reflects a locality principle in physics)  does not
fix  the renormalised integrals uniquely; even when the holomorphic
regularisation ${\cal R}$  is fixed (e.g. dimensional regularisation), there still remains  a freedom of choice
left due the freedom of choice on  the evaluator  unless one imposes further
constraints as one would do in quantum field theory.
 \\ 
This paper emphasises  the analytic  mechanisms underlying the renormalisation of
multiple  integrals of symbols with linear constraints, thereby raising further
analytic 
questions which remain to be solved, namely 
\begin{enumerate}
\item Do these results which hold for radial symbols extend to all
  classical symbols? The meromorphicity  established in  Theorem
  \ref{thm:meromultintsymb} easily extends to polynomial symbols when using Riesz
  or dimensional regularisation due to the fact that such symbols can be
  obtained from derivatives of radial symbols ($\xi_i= \frac{1}{2}\, \partial_i\vert \xi\vert^2$) but it is not clear whether one can go beyond
  those classes of symbols. 
\item   How do these renormalisation procedures generalise to integrals of tensor products of
  symbols with affine constraints so as to to allow for external momenta, one
  of the difficulties being how to control the symbolic behaviour of parameter dependent
  renormalised integrals in the  external parameters? 
\item How do the various renormalisation approaches described here compare?
 It follows  from the pole structure of the meromorphic extensions
 described in the paper that the renormalised values obtained by different methods
 coincide for symbols  $\sigma_i$ whose orders have non integer partial sums
 since the renormalised values  then correspond to ordinary
  evaluations of holomorphic maps at $0$, but it is not clear what happens
  beyond this case.
\item It would be interesting to investigate all the
  coefficients of the  Laurent expansion and to see when they can be
  recognized as motives \footnote{see \cite{BW2} and references therein for
    discussions along these lines}.
\end{enumerate}
Answering these questions can also be relevant for multiple discrete sums of symbols with
constraints (see \cite{P4}),  multiple zeta functions being an important instance since they
boil down to  mutiple  discrete sums of symbols with conical
constraints. 
\section*{Table of contents}
\subsection*{Part 1: Regularised integrals of symbols}
\subsubsection*{1. Cut-off regularised integrals of log-polyhomogeneous symbols}
\subsubsection*{2. Regularised integrals of log-polyhomogeneous symbols}
\subsubsection*{3. Basic properties of integrals of holomorphic  symbols}
\subsubsection*{4. Regularised integrals with affine parameters}
\subsection*{Part 2:  Renormalised multiple integrals of symbols with linear
  constraints}
\subsubsection*{5. Integrals of tensor products of   symbols revisited}
\subsubsection*{6. Linear constraints in terms of matrices}
\subsubsection*{7.  Multiple integrals of holomorphic families with constraints}
\subsubsection*{8. Renormalised integrals with constraints}
\vfill \eject \noindent
\section*{Part 1: Regularised integrals of symbols}
In this first part we review and partially extend results of \cite{MMP} and
\cite{MP}. Regularised integrals are defined using cut-off and holomorphic
regularisations: dimensional regularisation is presented as an instance of
holomorphic regularisations and then compared with cut-off regularisation.
\section{Cut-off regularised integrals of log-polyhomogeneous  symbols}
We recall    regularisation techniques for integrals of  
log-polyhomogeneous  symbols which deal with ultraviolet divergences. Starting from cut-off
regularisation we then turn to  dimensional
regularisation which we describe as an instance of more general holomorphic
regularisation procedures. We discuss in how far such regularisation procedures also
take care of   infrared divergences. Such issues were previously discussed by many
authors in the context of Feynman diagrams, starting with pioneering work of
t'Hooft and Veltman \cite{HV} on dimensional regularisation and later works of Smirnov 
\cite{Sm1}, \cite{Sm2},\cite{Sm3}   and Speer \cite{Sp} just to name a few
later developments.\\ Since integrating classical symbols naturally leads to
log-polyhomogeneous symbols, we describe regularisation procedures on the class
of log-polyhomogneous symbols. 
\subsection{From log-polyhomogeneous functions to  symbols }
We call a function $f\in \C^\infty(\R^n-\{0\})$
positively log-homogeneous of order $a$ and log-degree $k$ if\footnote{We use
  a terminology which is slightly different from that of \cite{L}}
$$f(\xi)= \sum_{l=0}^k f_{a, l}(\xi) \, \log^l \vert \xi\vert\quad f_{a,l}(t
\, \xi)=
t^{a}f_{a,l}(\xi)\quad \forall \xi\in \R^n, \quad \forall t>0.$$
Following \cite{L}, given a positively log-homogeneous function 
 of order $a$ and log degree $k$, let us  set for any $l\in \{1, \cdots, k\}$:
$$ {\rm res}_l (f):= \delta_{a+n}\, \int_{S^{n-1}}f_{-n, l}(\xi)\, \dbar_S\xi$$
where $d_S\xi$ is the volume form with respect to the standard metric on
$S^{n-1}$. Let us denote by ${\cal P}_+^{a, k}(\R^n)$ the set of positively
log-homogeneous functions on $\R^n$ of order $a$ and log degree $k$. 
\begin{ex}$\xi\mapsto f(\xi)= \sum_{l=0}^k c_{ l} \vert x\vert^a\, \log^l \vert \xi\vert$ with $c_{ l}\in \R, l=0, \cdots, k$  belongs to ${\cal P}_+^{a, k}(\R^n)$.
\end{ex}
We call a function $\sigma\in \Ci(\R^n)$ a log-polyhomogeneous  symbol of
order $a$ and log-type $k$    with constant
 coefficients if 
\begin{equation}\label{eq:localsymb}
\sigma= \sum_{j=0}^{N-1}\chi\, \sigma_{a-j}+ \sigma^\chi_{(N)},
\end{equation}
where $\chi$ is a smooth cut-off function which vanishes at $0$ and equals to one
outside the unit  ball, where  $ \sigma_{a-j}\in { \cal P}_+^{a-j,k}(\R^n)$ and where $\sigma^\chi_{(N)}\in \Ci(\R^n)$   satisfies the following requirement:
$$\exists C\in \R, \quad \vert \sigma  \chi_{(N)}(\xi)\vert \leq C \langle \xi\rangle^{{\rm Re}(a)-N}\quad \forall \xi\in \R^n$$
with $\langle \xi\rangle:= (1+\vert \xi\vert^2)^{\frac{1}{2}}$. Changing the
cut-off function $\chi$ amounts to modihying the remainder term  $\sigma^\chi_{(N)}$.\\ If the
log-type $k$ vanishes then the symbol is called polyhomogeneous or classical.\\ We call a symbol $\sigma$  radial if $\sigma(\xi)= f(\vert \xi\vert)$ only
depends on the radius.
\begin{rk} For short one writes $\sigma\sim \sum_{j=0}^\infty\chi\, \sigma_{a-j}$, the symbol $\sim$ controling the asymptotics as $\vert \xi\vert \to \infty$ i.e. the ultraviolet behaviour.
 \end{rk}
\begin{ex}$\sigma(\xi)= \frac{1}{\vert \xi\vert^2+1}$ is a classical radial symbol of
  order $-2$ and 
$$\sigma(\xi)\sim_{\vert \xi\vert \to \infty} \sum_{j=0}^\infty (-1)^k \vert \xi\vert^{-2-2k}.$$
\end{ex}
\begin{rk}\label{rk:infraredsymb} To deal with infrared divergences it can be
  useful to observe that a radial 
 function $$f(\xi)=\sum_{l=0}^k  c_l\, \vert\xi\vert^a\, \log^l \vert \xi\vert$$ in
 $ {\cal P}_+^{a, k}(\R^d)$ can be seen as a limit as $\e\to 0$ of radial symbols
 $$\sigma^\e (\xi)=\sum_{l=0}^k c_l \, (\vert \xi\vert^2+\e^2)^{\frac{a}{2}}\, \log^l
 \left((\vert \xi\vert^2+\e^2)^{\frac{1}{2}}\right).$$ When $\e\neq 0$ these
 are smooth functions on $\R^d$ which lie  in
 $CS^{a,k}(\R^d)$ and 
\begin{eqnarray*}
   \sigma^\e(\xi)&=&(1-\chi(\xi))\, \sigma^\e(\xi)+ \chi(\xi)\,\vert \xi\vert^{a}\,\sum_{l=0}^k   c_l \, \left(\left\vert
       \frac{\e}{\xi}\right\vert^2+1\right)^{\frac{a}{2}}\,\left(-a \log\vert \xi\vert
     +{\frac{1}{2}}\log \left(\left\vert
         \frac{\e}{\xi}\right\vert^2+1\right)\right)^l\\
&\sim& \sum_{j=0}^\infty \sigma^\e_{a-j} (\xi)\, \chi(\xi)\end{eqnarray*}
with $\sigma^\e_{ a-j}=\e^j \, \sigma^1_{a-j}$. As before, $\chi$ is a smooth cut-off function which vanishes in a
small neighborhood of $0$ and is one outside the unit ball. 
\end{rk} 
\begin{ex}Take $f(\xi)= \vert \xi\vert^{-2}$ which we write $f(\xi)=
  \lim_{\e\to 0}  (\vert \xi\vert^2+\e^2)^{-1}$. Then $$\sigma^\e(\xi)=
  \frac{1}{\vert \xi\vert^2+\e^2}\sim_{\vert \xi\vert \to \infty} \sum_{k=0}^\infty (-1)^k
  \vert\xi\vert^{-2-2k}\, \e^{2k}.$$
\end{ex}
Let $CS^{a, k}(\R^n)$ denote the set of log-polyhomogeneous symbols
with constant coefficients of order
$a$ and log-type $k$
\footnote{The following semi-norms labelled by multiindices $\gamma,\beta$ and integers $m\geq 0, p\in \{1, \cdots, k\}$, $N$ give rise to a Fr\'echet topology on
 $CS ^{a,k}( \R^n)$:
\begin{eqnarray*}
&{} & {\rm sup}_{\xi \in \R^n} (1+\vert \xi\vert)^{-a+\vert \beta\vert} \, \vert  \partial_\xi^\beta \sigma( \xi)\vert;\\
&{}&  {\rm sup}_{\xi\in \R^n}  \vert \xi\vert^{-a+N+\vert
  \beta\vert}\,\vert \partial_\xi^{\beta}
\left(\sigma-\sum_{m=0}^{N-1} \chi(\xi)\, \sigma_{a-m}\right)( \xi) \vert;\\
&{}& {\rm sup}_{ \vert\xi\vert=1}  
\vert  \partial_\xi^{\beta} \sigma_{a-m, p}( \xi) \vert.
\end{eqnarray*}}\label{footnote:Frechet}. It is convenient to introduce the following notation $CS^{*, k} (\R^n)= \cup_{a\in
\C}CS^{a, k}(\R^n)$. The algebra  $$CS(\R^n):= \langle \cup_{a\in
\C}CS^{a,0}(\R^n)\rangle$$ generated by all log-polyhomogeneous symbols of
log-type $0$ is called the algebra of classical or polyhomogeneous symbols on
$\R^n$ with constant coefficients.\\It contains the  algebra  $CS ^{-\infty}(\R^n):= \bigcap_{a\in \R} CS^{a}(\R^n)$  of smoothing symbols. 
The algebra  $$CS ^{\Z,*}(\R^n):= \bigcup_{a\in \Z}\bigcup_{k\in \N} CS ^{a,k}(\R^n)$$
of integer order log-polyhomogeneous symbols
\footnote{$CS ^{\Z,*}(\R^n)$ is equipped with an  inductive limit topology of Fr\'echet spaces} is
strictly contained in the algebra generated by log-polyhomogeneous symbols of
any order 
 $$CS ^{*,*}(\R^n):= \langle \bigcup_{a\in \C}\bigcup_{k\in \N} CS ^{a,k}(\R^n)\rangle.$$

\subsection{Cut-off regularised  integrals }
We recall the construction of cut-off regularised integrals of
log-polyhomogeneous symbols  \cite{L} which generalises results previously
established by Guillemin \cite{G} and Wodzicki \cite{W}  in the case of classical symbols.\\
For any non negative integer $k$ and any log-polyhomogeneous symbol $\sigma\in
CS ^{*,k}(\R^n)$, the expression $\int_{B(0, R)} \sigma(\xi) \dbar\xi$ has an asymptotic
expansion as $R$ tends to 
$\infty$ of  the form \footnote{We have set $\dbar\xi:= (2\pi)^{-n}\, d\xi$
  and  $\dbar\xi_i:= \frac{d\xi_i}{2\pi}$.}:
\begin{eqnarray}\label{eq:cutoffasympt}
\int_{B(0, R)} \sigma(\xi)  \dbar\xi&\sim_{ R\to \infty}&C(\sigma)+ \sum_{j=0,a-j+n\neq 0}^\infty \sum_{l=0}^k P_l(\sigma_{a-j, l})(\log R)  \, R^{a-j+n}\nonumber\\
&+&
\sum_{l=0}^k \frac{{\rm res}_{l}(\sigma)}{l+1}\log^{l+1}  R
\end{eqnarray}
where $${\rm res}_{l}(\sigma)= \int_{S^{n-1}} \sigma_{-n,l} (\xi)
\dbar_S\xi$$ is the higher $l$-th 
noncommutative residue, $P_l(\sigma_{a-j, l})(X)$ is a polynomial of
degree $l$ with coefficients depending on $\sigma_{a-j, l}$ and   $C(\sigma)$
is the constant term corresponding to the finite part  called the
{\it cut-off regularised integral of $\sigma$}:
\begin{eqnarray*}
\cutoffint_{\R^n} \sigma(\xi)\, \dbar\xi &
:=& \int_{\R^n}\sigma^\chi_{(N)}( \xi)\, d\xi+  \int_{B(0, 1)}
 \chi(\xi) \sigma( \xi)\, \dbar\xi\\
 &+&\sum_{j=0, a-j+n\neq 0}^{N-1}\sum_{l=0}^k \frac{(-1)^{l+1}l!}{(a-j+n)^{l+1}}
 \int_{S^{n-1}} \sigma_{a-j,l} (\xi) \dbar_S\xi\\
\end{eqnarray*}
with the notations of (\ref{eq:localsymb}).\\
It is independent of the choice of  $N\geq a +n-1$, as well as of the cut-off
function $\chi$. It is furthermore independent   of
the  parametrisation $R$ provided the higher noncommutative
residue  ${\rm res}_{l}(\sigma)$ vanish for all integer $0\leq l\leq k$ for we
have: 
$${\rm fp}_{R\to\infty}\int_{B(0,\mu \, R)}\sigma(\xi) d\xi={\rm
  fp}_{R\to\infty} \int_{B(0, R)} \sigma(\xi)\, d\xi+\sum_{l=0}^k
\frac{\log^{l+1}\mu}{l+1} \cdot{\rm res}_{l}(\sigma)$$
for any fixed $\mu>0$. \\

If $\sigma$ is a classical operator, setting $k=0$
in the above formula yields
\begin{eqnarray*}
\cutoffint_{\R^n}\sigma(\xi)\, \dbar\xi
 &:=& \int_{\R^n}\sigma_{(N)}( \xi) \, \dbar\xi+\sum_{j=0}^{N-1}  \int_{B(0, 1)}
 \chi(\xi)\, \sigma_{a-j}(\xi) \, \dbar\xi\\
&- &\sum_{j=0, a-j+n\neq 0}^{N-1}  \frac{ 1}{a-j+n}
 \int_{S^{n-1}} \sigma_{a-j } (\omega) \dbar_S\omega.
\end{eqnarray*}

\begin{rk}\label{rk:infraredreg} With the notations of Remark \ref{rk:infraredsymb}
we have the following Taylor expansion at $\e=0$:
\begin{eqnarray*}
\cutoffint_{\R^d}\sigma^\e(\xi)\, \dbar\xi
 &=& \int_{\R^n}\sigma^\e_{(N)}( \xi) \, d\xi+\sum_{j=0}^{N-1}  \int_{B(0, 1)}
 \chi(\xi)\, \sigma^\e_{ a-j}(\xi) \, \dbar\xi\\
&+& \sum_{j=0, a-j+n\neq 0}^{N-1}\sum_{l=0}^k \frac{(-1)^{l+1}l!}{(a-j+n)^{l+1}}\int_{S^{n-1}} \sigma^\e_{a-j,l}( \xi)\, \dbar_S\xi\\
&=&\sum_{j=0}^{N-1}  \e^j\, \int_{B(0, 1)}
 \chi(\xi)\, \sigma^1_{ a-j}(\xi) \, \dbar\xi\\
&+& \sum_{j=0, a-j+n\neq 0}^{N-1}\e^j\, \sum_{l=0}^k
\frac{(-1)^{l+1}l!}{(a-j+n)^{l+1}}\int_{S^{n-1}} \sigma^\e_{a-j,l}( \xi)\,  \dbar_S\xi\\
&+& O(\e^N)\\
\end{eqnarray*}
 since
$\sigma^\e_{(N)}= O(\e^N)$ as a result of the fact that $\sigma^\e\sim
\sum_{j=0}^\infty \e^j\, \sigma^\e_{a-j}$. It therefore
turns out that  the
regularised cut-off integral which is built  to  deal with ultraviolet divergences
 also naturally takes care of infrared divergences in as far as it yields a Taylor
expansion as $\e\to 0$ of the map $\e\mapsto \cutoffint_{\R^n} \sum_{l=0}^k
c_l \left(\vert \xi\vert^2+\e^2\right)^{\frac{a}{2}}\, \log^l
\left( \left(\vert \xi\vert^2+ \e^2\right)^\frac{1}{2} \right).$ 
\end{rk}
An important property of cut-off regularised integrals already observed in \cite{MMP} is that they vanish on
polynomials. 
\begin{prop}\label{prop:polynomials}
Let $P(\xi_1, \cdots, \xi_k)= \sum_{a} c_\alpha \xi_1^{\alpha_1
}\cdots \xi_k^{\alpha_k}$ be a polynomial expression in the $\xi_1, \cdots,
\xi_n$ with complex  coefficients $ c_\alpha$, then 
$$\cutoffint_{\R^n} P(\xi_1, \cdots, \xi_n)\, \dbar\,\xi =0$$
\end{prop}
{\bf Proof:} It suffices to prove that for any non negative integer $a$,
$$\cutoffint_{\R^n} \xi_i^\alpha d\,\xi=0.$$
Since for any $R>0$
\begin{eqnarray*}
\int_{B(0, R)} \xi_i^\alpha d\,\xi&=&\left( \int_0^R r^{n+\alpha-1}
 dr\right)\,\int_{S^{n-1}} \xi_i^\alpha d\,\xi\\
&=& \frac{R^{\alpha+n}}{\alpha+n} \,\int_{S^{n-1}} \xi_i^\alpha d\,\xi,\\
\end{eqnarray*}
we have 
$$\cutoffint_{\R^n} \xi_i^\alpha d\,\xi= {\rm fp}_{R\to
  \infty}\frac{R^{\alpha+n}}{\alpha+n} \,\int_{S^{n-1}} \xi_i^\alpha
d\,\xi=0. $$
\endsquare

\section{Regularised integrals of  log-polyhomogenous symbols}
\subsection{Cut-off regularised  integrals of holomorphic families}
Following \cite{KV} (see \cite{L} for the extension to log-polyhomgoeneous
symbols), we   call  a family $z\mapsto \sigma(z)\in CS^{*, k}(\R^n)$ of
logpolyhomogeneous symbols parametrised by $z\in\Omega\subset  \C$  holomorphic if the following
 assumptions hold:
 \begin{enumerate}
 \item the order $\alpha(z)$ of $\sigma(z)$
is holomorphic  in $z$,
\item for any $0\leq l\leq k$, for any non negative integer $j$, the
 homogeneous components
$\sigma_{ \alpha(z) -j, l} (z)$ of the symbol $\sigma( z) $
yield  holomorphic maps  into  $\Ci(\R^n )$,
\item   for any sufficiently large integer $N$, the map $$z\mapsto
  \int_{\R^n}e^{i\xi\cdot (x-y)}\, \left(\sigma(z)(\xi)- \sum_{j=0}^N
    \chi(\xi) \,\sigma_{\alpha(z)-j}(z)( \xi) \right)\, d\xi$$ 
 yields  a holomorphic map   $z\mapsto K^{(N)}$ into  some  $C^{K(N)} (
 \R^n\times \R^n)$  where $\lim_{N\to \infty} K(N)=+\infty$.
\end{enumerate}
We quote from \cite{PS}  the following theorem which extends results of
\cite{L} relating the noncommutative residue of holomorphic families of
log-polyhomogeneous symbols with higher  noncommutative residues. For simplicity, we
restrict ourselves to holomorphic  families with order $\alpha(z)$ given by an
affine function of $z$, a case which covers natural applications. 
 \begin{prop}\label{thm:KV} Let  $k$ be a non negative integer. For any holomorphic 
family  $z\mapsto\sigma(z)\in CS ^{\alpha(z), k}(\R^n)$ of
symbols parametrised by  a domain  $W\subset \C$
such that $z\mapsto \alpha(z)=\alpha^\prime(0)\, z+ \alpha(0)$ is  a non
constant  affine
function, there is a Laurent expansion in a neighborhood of any $z_0\in \C$
\begin{eqnarray*}
\cutoffint_{\R^n}
\sigma(z)(
\xi) \dbar\xi &=& {\rm fp}_{z=z_0}\cutoffint_{\R^n}
\sigma(z)(
\xi) d\xi
+ \sum_{j=1}^{k+1} \frac{r_j(\sigma)(z_0) (x)}{(z-z_0)^{j}} \\
&+& \sum_{j=1}^{K} s_j(\sigma)(z_0) (x)\, (z-z_0)^{j}\\
 + o\left((z-z_0)^K\right),\\
\end{eqnarray*}
 where for $1\leq j\leq k+1$,
$r_j(\sigma)(z_0) (x)$ is   explicitly determined by a local expression (see \cite{L} for the case $\alpha^\prime(0)=1$)
\begin{equation} \label{eq:KVlogsymbresj} r_j(\sigma)(z_0) (x)
:=\sum_{l=j-1}^k \frac{(-1)^{l+1}}{(\alpha^\prime(z_0))^{l+1}}\frac{l!}{(l+1-j)!}\,  \,{\rm res} \left( \left(\sigma_{( l)}\right)^{(l+1-j)}\right) (z_0).
\end{equation}
Here ${\rm res}(\tau)= \int_{S^{n-1}} \tau_{-n,0}(\xi)\, \dbar_S\xi$, $\sigma_{(l)}(z)$ is the local symbol given by the coefficient of $\log^l \vert \xi\vert$ of $\sigma$ i.e. $\sigma(z)= \sum_{l=0}^k \sigma_{(l)}(z)\log^l \vert \xi\vert.$
On the other hand,  the finite part ${\rm
  fp}_{z=z_0}\cutoffint_{\R^n}\sigma(z)(\xi) \dbar\xi $ consists of a global piece
given by the cut-off regularised integral $\cutoffint_{\R^n} \sigma(z_0)(
\xi)\, \dbar\xi$ and a local piece expressed in terms of residues:
\begin{eqnarray} \label{eq:KVlogsymb0}
{\rm fp}_{z=z_0}\cutoffint_{\R^n}
\sigma(z)(
\xi) \dbar\xi& = &\cutoffint_{T_x* U} \sigma(z_0)( \xi)\, \dbar\xi \nonumber\\
&+& \sum_{l=0}^k \frac{(-1)^{l+1}}{(\alpha^\prime(z_0))^{l+1}}\frac{1}{l+1}\,  \,{\rm res} \left( \left(\sigma_{( l)}\right)^{(l+1)}\right) (z_0).
\end{eqnarray}
\end{prop}
As a consequence, the finite part ${\rm fp}_{z=z_0}\cutoffint_{\R^n}
\sigma(z)(
\xi) d\xi$ is entirely determined by the derivative $\alpha^\prime(z_0)$ of
the order and by the derivatives of the symbol  $\sigma^{(l)}(z_0), \quad
l\leq k+1$  via the cut-off  integral and the noncommutative residue. 

\subsection{Regularised integrals }
    Let us briefly recall the  notion of holomorphic regularisation taken
    from \cite{KV} (see also \cite{PS}) and adapted to physics applications
    in \cite{P1}.     It includes dimensional regularisation used in
    perturbative quantum  field theories to cure singularities arising in loop
    diagrams see e.g. \cite{HV}, \cite{Sm1}, \cite{Sm2}, \cite{Sm3}.                             
\begin{defn} 
A  holomorphic regularisation procedure on a subset ${\cal S}\subset CS ^{*, *}(\R^n)$ 
is a  map 
$\sigma  \mapsto  (z\mapsto \sigma(z))$ which sends $\sigma\in  CS ^{*, k}(\R^n)$ to a holomorphic family 
$\sigma\in CS ^{*, k}(\R^n)$ 
such that 
\begin{enumerate}
\item  $\sigma(0)=\sigma$, 
\item  $\sigma(z)$ has holomorphic  order  $\alpha(z)$
 (in particular, $\alpha(0)$ is equal to the order of $\sigma$) 
  such that $\alpha^\prime(0)\neq 0$.\\
 \end{enumerate}
We call a regularisation procedure  ${\cal R}$ continuous whenever   the map 
$
\sigma \mapsto\left( z\mapsto \sigma(z)\right)$
  is continuous for the Fr\'echet topology on $CS^{a, k}(\R^n)$ (see previous  footnote). 
\end{defn}
 
One often comes across holomorphic regularisations of the type:
$${\cal R}(\sigma)(z)= \sigma\cdot \tau(z)$$ 
 where  $\tau(z)$ is a
holomorphic family of symbols in $CS ( \R^n)$ such that
\begin{enumerate}
\item $\tau(0)=1$,
 \item $ \tau(z)$ has holomorphic order $-q\, z$ with $q>0$.
\end{enumerate} Note that this implies  that $\sigma(z)$ has order $\alpha(z)= \alpha(0)- q\,
z$ with $q\neq 0$. \\
This class of holomorphic regularisations contains known regularisation such as
\begin{itemize}
\item  Riesz
regularisation for which $\tau(z)(\xi):=\chi(\xi)\,   \vert
\xi \vert^{-z}$, where $\chi $ is some smooth cut-off function around $0$ which
is equal to $1$ outside the unit ball. 
\item This is a particular instance of regularisations for which  
   $\tau(z)(\xi)=H(z)\cdot \chi(\xi)\,  \vert \xi \vert^{-z}$ where $H$ is a
holomorphic function such that $H(0)=1$.
\item In even dimensions, dimensional regularisation corresponds to the choice (see \cite{P1})
$H(z):=\frac{\pi^{-\frac{z}{2}}
  \Gamma\left(\frac{n}{2}\right)}{\Gamma\left(\frac{n-z}{2}\right)}$ which is
a holomorphic function at $z=0$ such that $H(0)=1$.
\end{itemize}
The function
$H$ stands for the relative volume of the unit cotangent sphere in dimension
$n$ w.r.to its ``volume'' in dimension $n(z)= n-z$.
\begin{ex}  To illustrate this, let us
consider integrals of a radial symbol
$\sigma(\xi):= f(\vert \xi\vert)$  following the physicists' prescription for
dimensional regularisation. Assuming
that the symbol has order with real part  $<-n$, then  $$ \int_{\R^n}\sigma(\xi)\, d^{n}\xi=
{\rm Vol}(S^{n-1})\, \int_{\R^n}f(r)\, r^{n-1} dr=\frac{2\,\pi^{p}}{ \Gamma\left(p\right)}\, \int_{\R^n}f(r)\, r^{n-1} dr$$ 
 since the volume of the unit sphere $S^{n-1}$ in even dimensions $n=2p$  is given by 
$
{\rm Vol} (S^{n-1})=\frac{2 \pi^k } {( k-1)!}= \frac{2\,\pi^{p}}{
  \Gamma\left(p\right)}$. Replacing $n$ by $n-z$ in the above
expression yields a holomorphic map on the half plane ${\rm Re}(z)<{\rm Re}(a)+n$: 
$$z\mapsto \frac{2\,\pi^{p-\frac{z}{2}}}{ \Gamma\left(p-\frac{z}{2}\right)}\,
\int_{\R^n}f(r)\, r^{n-z-1} dr= \int_{\R^n}\sigma(z)(\xi)\, \dbar\xi $$ 
where we have set $\sigma(z)(\xi)=H(z)\,  \sigma(\xi)\,\vert \xi\vert^{-z} $
 and $H(z):=
 \frac{2\,\pi^{p-\frac{z}{2}}\Gamma\left(p\right)}{2\,\pi^{p} \Gamma\left(p-\frac{z}{2}\right)}= 
 \frac{\pi^{-\frac{z}{2}}\Gamma\left(p\right)}{
   \Gamma\left(p-\frac{z}{2}\right)}$. By the above constructions, we know
 that this extends to a meromorphic map 
$z\mapsto \cutoffint_{\R^n}\sigma(z)(\xi)\, dx $ on the  whole complex plane.
Thus,  dimensional regularisation on
 radial symbols boils down to    holomorphic regularisation on the integrand.
\end{ex}
These regularisation procedures are clearly
continuous. They have in common that the order $\alpha(z)$ of $\sigma(z)$ is
affine in $z$:
\begin{equation}\label{eq:affineorder}
\alpha(z)= \alpha(0)-q\, z, \quad q \neq
0,\end{equation} which is why we restrict to this situation.  \\ \\ As a consequence of the results of the previous paragraph, given
a holomorphic regularisation procedure ${\cal R}:\sigma \mapsto \sigma(z)$ on
$CS^{*, k}(\R^n)$  and a symbol $\sigma\in CS^{*, k}(\R^n)$, the map $z\mapsto \cutoffint_{\R^n} \,  \sigma(z)(\xi) \, d\xi$  is
meromorphic with  poles of order at most $k+1$ at points in $\alpha^{-1}([-n,
+\infty[\, \cap\, \Z)$ where $\alpha(z)$ is the order of $\sigma(z)$ so that we can define the finite part when $z\to 0$   as follows. 
\begin{defn}
Given  a holomorphic regularisation procedure ${\cal R}:\sigma \mapsto
\sigma(z)$ on $CS ^{*, k}(\R^n)$ and a symbol 
$\sigma\in CS ^{*, k}(\R^n)$,  we define the regularised integral 
\begin{eqnarray*}
\cutoffint_{\R^n}^{{\cal R}} \sigma( \xi) \, \dbar\xi &:=& {\rm fp}_{z=0} \cutoffint_{\R^n} \sigma(z)( \xi)\, \dbar\xi\\
&:=& \lim_{z\to 0}\left(  \cutoffint_{\R^n} \sigma(z)( \xi)\,\dbar\xi \, - \sum_{j=1}^{k+1}
\frac{1}{z^j} {\rm Res}^j_{z=0}\cutoffint_{\R^n}  \sigma(z)(
\xi)\,\dbar\xi\,\right).
\end{eqnarray*}
In particular, in even dimensions    we  define the dimensional regularised
integral of $\sigma$ by
\begin{equation}\label{eq:dimreg}\cutoffint_{\R^n}^{\rm dim.reg}\sigma:= {\rm fp}_{z=0}\left( H(z)\cutoffint_{\R^n}\chi(\xi)\,
\sigma(\xi)\, \vert \xi\vert^{-z}\, \dbar \xi\right)+\int_{\R^n}(1-\chi(\xi))
\sigma(\xi)\, d\xi
\end{equation}
which is independent of the choice of cut-off function $\chi$.
\end{defn}
\begin{ex}\label{ex:Riesz} Simple computations show  that Riesz  and cut-off
  regularised integrals of symbols coincide.
\end{ex}
\begin{thm} \label{thm:cutoffdimreg}Dimensional regularised integrals of symbols in $CS^{*, k}(\R^n)$
  with $n=2p$ even differ from cut-off regularised
  integrals by  a linear combination of the first $k+1$ derivatives of the function  $H(z):=\frac{\pi^{-\frac{z}{2}}
  \Gamma\left(p\right)}{\Gamma\left(p-\frac{z}{2}\right)}$ with
coefficients
involving the residues of derivatives of the symbol:
\begin{equation}
 \cutoffint^{\rm dim.reg}_{\R^n}  \sigma(\xi) \dbar \, \xi=
\cutoffint_{\R^n}\sigma(\xi) d \, \xi+\sum_{l=0}^kH^{(l+1)}(0)\,\sum_{j=l}^{k} \frac{j!}{(j-l)!}\,  \,{\rm res} \left( \left(\sigma_{( j)}\right)^{(j-l)} (0)\right).
\end{equation} When 
$k=0$,  $\sigma$ is classical and: $$
 \cutoffint^{\rm dim.reg}_{\R^n}\sigma(\xi) \dbar \, \xi=  \cutoffint_{\R^n}\sigma(\xi) \dbar \, \xi 
+\frac{1}{2}\left( \left(\sum_{j=1}^{p-1} \frac{1}{j}
+\gamma\right) -\log \pi \right)\cdot {\rm res}(\sigma).$$
\end{thm}
{\bf Proof:} The fact that   
dimensional regularisation is obtained from Riesz regularisation
$\sigma\mapsto \sigma(z)$ by   multiplying $\sigma(z)$ by a
  function $H(z)$ introduces  extra terms   involving  complex residues:
$$ {\rm fp}_{z=0}\left(H(z)
 \cdot  \cutoffint_{\R^n}  \sigma(z)(\xi) d \, \xi\right)=
\cutoffint_{\R^n}\sigma(\xi) d \, \xi+ \sum_{l=0}^k {\rm Res}^{l+1}(\sigma(z))
H^{(l+1)}(0)$$ which, when combined with 
 (\ref{eq:KVlogsymbresj})  yields:
\begin{eqnarray*}
{\rm fp}_{z=0}\left(H(z)
 \cdot  \cutoffint_{\R^n}  \sigma(z)(\xi) \dbar \, \xi\right)&=& 
\cutoffint_{\R^n}\sigma(\xi) d \, \xi\\
&+&\sum_{l=0}^kH^{(l+1)}(0)\,\sum_{j=l}^{k} \frac{j!}{(j-l)!}\,  \,{\rm res} \left( \left(\sigma_{( j)}\right)^{(j-l)} (0)\right)
\end{eqnarray*}
since $\alpha(z)= \alpha(0)-z$. 
In particular, when $\sigma$ is classical (i.e. when  $k=0$) we have:
$$ {\rm fp}_{z=0}\left(H(z)
 \cdot   \cutoffint_{\R^n} \sigma(z)(\xi) \dbar \, \xi\right)=
   \cutoffint_{\R^n}\sigma(\xi) d \, \xi+  {\rm res} (\sigma) \cdot
  H^\prime(0).$$ Since derivatives  at $p\in \N-\{1\}$ of the Gamma function read:
$\Gamma^\prime\left(p\right)=
\Gamma(k)\left(\sum_{j=1}^{p-1}\frac{1}{j}-\gamma\right)$ it follows that 
$$H^\prime(0)=\frac{1}{2}\left( -\log \pi +
\frac{\Gamma^\prime(p)}{\Gamma(p)}\right)= \frac{1}{2}\left(
-\log\pi+ \left(\sum_{j=1}^{p-1} \frac{1}{j}
+\gamma\right)\right).$$The result then follows.
\endsquare
\begin{rk} Since we saw in Remark   \ref{rk:infraredreg} that cut-off regularisation takes care of infrared
  divergences as well as ultraviolet ones, it follows that so does dimensional
  regularisation take care of infrared divergences.  The additional residue
  terms only contribute by additional terms in the Taylor expansion at $\e=0$.
\end{rk}
To close this paragraph,  we observe  that just as  the cut-off regularised integral was,
the map $
\sigma\mapsto \cutoffint_{\R^n}^{{\cal R}}\sigma( \xi) \, \dbar\xi$
is continuous for any continuous holomorphic regularisation procedure ${\cal R}:\sigma
\mapsto \sigma(z)$ on $CS ^{*, k}(\R^n)$, $k\in \N$.
\section{Basic properties of  integrals of holomorphic symbols}
Cut-off regularisation turns out to have nice properties for non integer order
symbols, such as a Stokes' property, translation invariance and covariance.
Consequently, computations involving dimensional regularisation can be carried
out following the usual integration rules such as integration by parts, change of
variables as long as this is done before taking finite parts. This in fact
holds for any holomorphic regularisation procedure as is shown below, so in
particular for dimensional regularisation, and 
provides a mathematical justification for the computations carried out by
physicists when performing changes of variable and integrations by parts.
\subsection{Integration by parts}
An important property of cut-off regularised integrals is
integration by parts, which is an instance of a more general Stokes' property
for symbol valued forms investigated in \cite{MMP}. 
\begin{prop} \label{prop:Stokes}
Let $\sigma\in CS^{*,*} (\R^n)$ with order $\alpha\notin \Z\cap[-n,
\infty[$. Then for any multiindex $\alpha$, 
$$\cutoffint_{\R^n}  \partial^\alpha \sigma( \xi)\,d\,\xi = 0.$$
\end{prop}
\begin{rk} This Stokes' property actually characterises
  the cut-off regularised integral $\cutoffint_{\R^n}$ in as far as it is the
  only linear extension of the ordinary integral to  non integer order symbols
  which vanishes on derivatives \cite{P2}.
\end{rk}
{\bf Proof:} We recall the general lines of the  proof and refer to \cite{MMP}
for further details. We only prove the result for classical symbols since the
proof easily extends to log-polyhomogeous symbols. It is clearly sufficient to
prove the result for a multiindex $\alpha=i$ of length one.
\begin{itemize}
\item 
If $\sigma$ has order $<-n$ then  we write:
\begin{eqnarray*}
\cutoffint_{\R^n}  \partial_{\xi_i} \sigma(
\xi)\, d\,\xi&=&\int_{\R^n}  \partial_{\xi_i} \sigma( \xi)\, d\,\xi\\
&=& \lim_{R\to 0} \int_{B(0, R)}d\,\xi\,  \partial_{\xi_i} \sigma(
\xi)\\
&=&(-1)^{i-1} \lim_{R\to 0} \int_{B(0, R)}  d\left( \sigma(
\xi)\, \dbar\xi_1\wedge \cdots d\hat \xi_i\wedge \cdots \wedge \dbar\xi_n\right)\\
&=&(-1)^{i-1} \lim_{R\to 0} \int_{S(0, R)} \sigma(
\xi)\, \dbar_S\xi\\
\end{eqnarray*}
where we have set  $d_S\xi:=\dbar\xi_1\wedge \cdots \wedge d\hat \xi_i\wedge \cdots \wedge 
\dbar\xi_n$.\\
This limit vanishes. Indeed, $\sigma$ being a symbol of order $\alpha$, there is
a positive constant $C$ such that
\begin{eqnarray*}
\vert \int_{S(0, R)}  \sigma(
\xi)\vert\, d_S\, \xi\, &\leq &\int_{S(0, R)}\vert  \sigma(
\xi)\vert\, d_S\xi \\
& \leq & C\int_{S(0, R)} (1+\vert \xi\vert^2)^{\frac{\alpha}{2}}\,d_S\xi\,\\
&\leq &C\,  R^n\, (1+ R^2)^{\frac{\alpha}{2}}\, {\rm Vol} \left(S^{n-1}\right).
\end{eqnarray*}
Here $S(0, R)\subset B(0, R)$ is the sphere of radius $R$ centered at
$0$  in  $\R^n$. 
\item
If $\alpha\geq -n$, we write $\sigma= \sum_{j=0}^N \chi(\xi)\,
\sigma_{\alpha-j}( \xi)+ \sigma_{(N)}( \xi)$
where $\chi$ is a smooth cut-off function, $\sigma_{\alpha-j}( \xi)$ is
positively homogeneous of degree $\alpha-j$ and $N$ is chosen large enough for
$\sigma_{(N)}$ to be  of order
$<-n$.  We have:
$$\cutoffint_{\R^n} \partial_{\xi_i} \sigma(
\xi)\,d\,\xi= \sum_{j=0}^N \cutoffint_{\R^n}\chi(\xi)  \partial_{\xi_i} \sigma_{\alpha-j}(
\xi) \, d\, \xi+ \int_{\R^n}  \partial_{\xi_i} \sigma_{(N)}(
\xi)\, d\, \xi\,.$$
 It follows from the above computation that $\int_{\R^n}  \partial_{\xi_i} \sigma_{(N)}(
\xi)\, d\, \xi=0$. On the other hand, we have for large enough $R$ and any positive
integer $i$: 
\begin{eqnarray*}
&{}&\cutoffint_{\R^n} \partial_{\xi_i} \left(\chi(\xi)\sigma_{\alpha-j}(
\xi)\right)\,d\,\xi\\
&=& {\rm fp}_{R\to \infty} \left(\int_{B(0, R)} \partial_{\xi_i} \left(\chi(\xi) \,\sigma_{\alpha-j}(
\xi)\right)\,d\,\xi\right)\\
&=& (-1)^{i-1} {\rm fp}_{R\to \infty}\left( \int_{S(0, R)}\chi(\xi)\, \sigma_{\alpha-j}(
\xi)\, d_S\, \xi\right)\\
&=& {\rm fp}_{R\to \infty} \left(\int_{S(0, R)}  \sigma_{\alpha-j}(
\xi)\,d_S\, \xi \right)\\
&{\rm since}& \quad \chi_{ \vert_{ S(0,R)}}=1\quad {\rm for}\quad {\rm
  large}\quad R\\
&=& {\rm fp}_{R\to \infty}\left( R^{\alpha-i-j+n+1} \int_{S^{n-1}} \sigma_{\alpha-j}(
\xi)\,d_S\, \xi\right)\\
\end{eqnarray*}
which vanishes if $\alpha+n \notin \N\cup \{0\}$.
\end{itemize}
\endsquare \\ \\
Let as before  ${\cal R}:\sigma\mapsto \sigma(z)$
be a holomorphic regularisation on $CS (\R^n)$. The following result is a
direct consequence of the above proposition. 
\begin{cor}\label{cor:holintegrationbyparts}\cite{MMP}
The
following equality of meromorphic functions holds:  
$$\cutoffint_{\R^n} \partial^\alpha (\sigma(z))( \xi)\,\dbar\,\xi\, =0$$
for any  multiindex $\alpha$  and any $\sigma\in CS^{*,*} (\R^n)$.
 
\end{cor}
{\bf Proof:}  The maps $z\mapsto  
\cutoffint_{\R^n}  \partial^\alpha \sigma(z)( \xi)\, d\,\xi$ are meromorphic
as  cut-off regularised integrals of a holomorphic family of symbols with poles in
$\alpha^{-1}\left(\Z\cap [-n ,+\infty[\right)$. 
By Proposition \ref{prop:Stokes} the expression  $\cutoffint_{\R^n}  \partial^\alpha
(\sigma(z))( \xi)\, d\,\xi\,$ vanishes outside these poles so that the identity announced in the
corollary   holds as an
equality of meromorphic maps. 
\endsquare
\begin{rk} This does not imply that the same properties hold for
  $\cutoffint^{\cal R}$. Unless the total order of the symbols is non integer,
  one is in general to expect that
$$\cutoffint_{\R^n}^{\cal R}  \partial_\xi^i \sigma( \xi) d\,\xi\,\neq 0.$$
\end{rk}
\subsection{Translation invariance}
We let $\R^n$ act  on
$CS ^{*,*}(\R^n)$ by translations as follows\footnote{$\eta$  can be seen  as external
momentum (usually denoted by $p$) whereas $\xi$ plays the role of internal
momentum (usually denoted by $k$) in physics}:
\begin{eqnarray*}
\R^n\times CS ^{*,*}(\R^n)&\to & CS ^{*,*}(\R^n)\\
(\eta, \sigma)&\mapsto& t_\eta^*\sigma( \xi):= \sigma( \xi+\eta).
\end{eqnarray*}
The map $\eta \mapsto t_{ \eta}^* \sigma:= \sigma(\cdot+  \eta)$ has the following Taylor expansion
  at $\eta=0$: 
\begin{equation}\label{eq:Taylorexpansion}t_{\eta}^* \sigma(\xi):= \sum_{\vert \beta\vert\leq N}\partial^\beta \sigma (\xi)\,
\frac{\eta^\beta}{\beta!} +\sum_{\vert \beta\vert=N+1}\frac{
\eta^\beta }{\beta!} \,
 \, \int_0^1 (1-t)^N\, \partial^\beta  \sigma (\xi+ t\eta)\,  dt\quad
\forall \xi\in \R^n.
\end{equation} 
Let us recall the following translation property for non integer order symbols.
\begin{prop}\label{prop:translinv}\cite{MP}
For any  $\sigma \in  CS^{*,*} (\R^n)$ and any $\eta\in \R^n$, the cut-off
integral 
$$\cutoffint_{\R^n}\sigma( \xi+\eta)\, \dbar\xi:= {\rm fp}_{R\to \infty}
\int_{B(0, R)}\sigma( \xi+\eta)\, \dbar\xi$$
is well defined. If $\sigma$ has order 
  $a\notin \Z\cap [-n,
+\infty[$ then: 
$$\cutoffint_{\R^n}t_\eta^*\sigma( \xi)\, \dbar\xi= \cutoffint_{\R^n}\sigma( \xi)\, \dbar\xi.$$
\end{prop} 
\begin{rk} Translation invariance actually characterises
  the cut-off regularised integral $\cutoffint_{\R^n}$ in as far as it is the
  only translation invariant linear extension of the ordinary integral to  non
  integer order symbols \cite{P2}, \cite{P3}.
\end{rk}
{\bf Proof:} Let $\sigma \in CS^{a,*}(\R^n)$.
\begin{itemize}
\item If $a<-n$ then $$\cutoffint_{\R^n}t_\eta^*\sigma( \xi)\, \dbar\xi=\lim_{R\to \infty}
\int_{B(0, R)}\sigma( \xi+\eta)\, \dbar\xi=
  \int_{\R^n}\sigma( \xi)\, \dbar\xi$$ is well defined.  The second part of
  the statement  then  follows from
  translation invariance of the ordinary Lebesgue integral. 
\item Let us assume ${\rm Re}(a)\geq -n$. The derivatives
$\partial^\beta\sigma$ arising in the Taylor
expansion 
  (\ref{eq:Taylorexpansion}) lie in  $CS (\R^n)$ so that their integrals over the ball
$B(0, R)$  have asymptotic expansions when $R\to \infty$ in decreasing powers of $R$ with a finite number of
powers of $\log R$. For  $\vert \beta\vert= N+1$ with $N$ chosen large enough, the
asymptotic expansion converges as $R$ tends to infinity and has no logarithmic term.  The integral $ \int_{B(0, R)}\sigma( \xi+\eta)\, \dbar\xi$
therefore  has the same type of asymptotic expansion when $R\to \infty$ as  $ \int_{B(0, R)}\sigma( \xi)\, \dbar\xi$
and  the finite part : 
\begin{eqnarray*}
\cutoffint_{\R^n}\sigma( \xi+\eta)\, \dbar\xi&:=& {\rm fp}_{R\to \infty}
\int_{B(0, R)}\sigma( \xi+\eta)\, \dbar\xi\\
&=&
\sum_{\vert \beta\vert\leq
  N}\cutoffint_{\R^n} \partial^\beta \sigma \,
\frac{\eta^\beta}{\beta!} +\sum_{\vert \beta\vert=N+1}\frac{
\eta^\beta }{\beta!} \,
 \, \int_0^1 (1-t)^N\,  \int_{\R^n}\partial^\gamma  \sigma (\cdot+ t\eta)\,  dt
\\
\end{eqnarray*}
is well defined.\\
Mimicking the proof of  Proposition \ref{prop:Stokes}, for $\vert
\beta\vert>0$ we write $\partial^\beta=
\partial_{\xi_i}\circ\partial^\gamma$ for some index $i $ and some multiindex $\gamma$:
 \begin{eqnarray*}
\cutoffint_{\R^n} \, \partial^{\beta}\sigma(
  \xi+\theta\,\eta)\, \dbar\xi&=& \int_{\R^n} \,\partial_{\xi_i}\left( \partial^{\gamma}\sigma(
  \xi+\theta\,\eta)\right)\, \dbar\xi\\
&=&\lim_{R\to  \infty} \int_{B(0, R)} \,\partial_{\xi_i}\left( \partial^{\gamma}\sigma(
  \xi+\theta\,\eta)\right) \, \dbar\xi\\
&=&(-1)^{i-1} \lim_{R\to  \infty} \int_{S(0, R)} \, \partial^{\beta}\sigma(
  \xi+\theta\,\eta)\, d_S\xi \\
\end{eqnarray*}
where as before we have set $d_S\xi:= (-1)^{i-1}\, \dbar\xi_1\wedge
\cdots\wedge d\hat \xi_i\wedge\cdots \wedge \dbar\xi_n$.
Since  $\sigma$ and hence its derivatives are
symbols, there is a positive constant $C$ such that for $\vert\beta\vert= N+1$
chosen 
large enough we have
\begin{eqnarray*}
\vert \int_{S(0, R)} \, \partial^{\gamma}\sigma(
  \xi+\theta\,\eta)\vert &\leq &  \int_{S(0, R)} \,\vert  \partial_\xi^{\gamma}\sigma(
  \xi+\theta\,\eta)\vert \\
&\leq & C \int_{S(0, R)} \,(1+ 
 \vert \xi+\theta\,\eta\vert)^{{\rm Re}(a)-(N+1)} \\
&\leq & C \, {\rm Vol}(S^{n-1}) \,R^n\, (1+ 
 \vert R- \vert \theta\,\eta\vert\vert)^{{\rm Re}(a)-(N+1)} \\
\end{eqnarray*}
which tends to $0$ as $R\to \infty$. Hence the cut-off regularised integral of the
remainder term vanishes. \\
If moreover $a\notin \Z$ then by Proposition \ref{prop:Stokes}, we have
$\cutoffint_{\R^n} \,\partial^\beta \sigma( \xi)=0$ for any  non vanishing
multiindex $\beta$.
Hence, only the $\beta=0$ term remains in the Taylor expansion and the result
follows. 
\end{itemize}
\endsquare\\ \\
Let as before  ${\cal R}:\sigma\mapsto \sigma(z)$
be a holomorphic regularisation on $CS (\R^n)$. The following result is a
direct consequence of the above proposition. 
\begin{cor}\label{cor:holtranslinv}\cite{MMP}
For any  $\sigma\in CS ^{*,*}(\R^n)$ and  any  $\eta\in \R^n$ the
following equality of meromorphic functions holds:  
$$\cutoffint_{\R^n}  \sigma(z)( \xi+\eta) (
\xi)\,d\, \xi= \cutoffint_{\R^n}  \sigma(z)( \xi)\,d\, \xi.$$
\end{cor}
{\bf Proof:} The Taylor expansion
\begin{eqnarray*}
&{}&\cutoffint_{\R^n}\sigma(z)(\xi+\eta)\, \dbar\xi\\
&=&\sum_{\vert \beta\vert\leq
  N}\cutoffint_{\R^n} \partial^\beta \sigma(z) \,
\frac{\eta^\beta}{\beta!} +\sum_{\vert \beta\vert=N+1}\frac{
\eta^\beta }{\beta!} \,
 \, \int_0^1 (1-t)^N\,  \int_{\R^n}\partial^\gamma  \sigma(z) (\cdot+ t\eta)\, 
 dt
\end{eqnarray*}
provides  meromorphicity of  the map  $z\mapsto
\cutoffint_{\R^n}\sigma(z)( \xi+\eta)\, \dbar\xi$ since we know that the maps  $z\mapsto 
\cutoffint_{\R^n}  \partial_\xi^\beta\sigma(z)( \xi)$ are meromorphic as
cut-off regularised integrals  of holomorphic families of ordinary symbols  and since the map
given by the remainder term  $z\mapsto \sum_{\vert \beta\vert=N+1}\int_{\R^n}\partial^{\beta}\sigma(z)(
  \xi+\theta\,\eta)\, d \xi$ is holomorphic for large enough $N$. Outside the set of poles we have by 
Proposition  \ref{prop:translinv} that $\cutoffint_{\R^n}\sigma(z)(
\xi+\eta)\, d\,\xi= \cutoffint_{\R^n}\sigma(z)( \xi)\, d\,\xi$ so that
the equality holds as an equality of  meromorphic functions.
\endsquare\\ 

\begin{rk} This does not imply that translation invariance  holds for
  $\cutoffint^{\cal R}$. Unless the  order of the symbol is non integer,
  one is in general to expect that
$$\cutoffint_{\R^n}^{\cal R}d\,\xi\, \sigma( \xi+\eta) \neq  \cutoffint_{\R^n}^{\cal R} d\,\xi \, \sigma( \xi) .$$
\end{rk}
\subsection{Covariance}
 $GL_n(\R^n)$ acts on $CS^{*, *} (\R^n)$
as follows 

\begin{eqnarray*}
GL_n(\R^n)\times CS ^{*,*}(\R^n)&\to & CS ^{*,*}(\R^n)\\
(C, \sigma)&\mapsto & \left(  \xi\mapsto \sigma( C\, \xi)\right).
\end{eqnarray*}

We quote  from \cite{L} the following extension  to
log-polyhomogeneous symbols of a result proved in
\cite{KV} for classical symbols. 
\begin{prop}\label{prop:covariance}
Let $\sigma\in CS^{a,*} (\R^n)$ with order $a\notin
\Z\cap[-n, \infty[$. Then for any $C\in GL_n(\R^n)$
$$\vert {\rm det}\, C\vert\, \cutoffint_{\R^n} \sigma( C\, \xi) \, d\,\xi=
\cutoffint_{\R^n} \sigma( \xi)\,  d\,\xi.$$
\end{prop}
 Let as before  ${\cal R}:\sigma\mapsto \sigma(z)$
be a holomorphic regularisation on $CS (\R^n)$. The following result is a
direct consequence of the above proposition. 
\begin{cor}\label{cor:holcovariance}
For any  $\sigma\in CS ^{*,*}(\R^n)$ and   for any $C\in GL_n(\R^n)$ the
following equality of meromorphic functions holds:  
$$\vert {\rm det} \, C\vert \, \cutoffint_{\R^n}   \sigma(z)( C\, \xi)\,
d\,\xi\,= \cutoffint_{\R^n} \sigma(z)( \xi)\, d\,\xi \, .$$
\end{cor}
\begin{rk} This does not imply the covariance of the regularised integral
  $\cutoffint^{\cal R}$. Unless the  order of the symbols is non integer,
  one is in general to expect that
$$\vert {\rm det} \, C\vert \, \cutoffint_{\R^n}^{\cal R}  \sigma(C\,  \xi)\,d\,\xi\, \neq  \cutoffint_{\R^n}^{\cal R} \sigma( \xi)\,d\,\xi \, .$$
\end{rk}
  \section{Regularised integrals with affine parameters}
The aim of this section is to regularise and then investigate the dependence in the
  external parameters $p_i$ of a priori divergent integrals of the type  
\begin{equation}\label{eq:Feynmanint}\int_{\R^n} \frac{P(k, p_1,\cdots, p_J)}{\left(\left(L_1(k, p_1, \cdots,p_J)\right)^2+
  m^2\right)^{s_1}\cdots\left( \left(L_I(k, p_1, \cdots,p_J)\right)^2+m^2\right)^{s_I}}\,
dk,
\end{equation}
where $P(k, p_1, \cdots, p_J)$ is a polynomial expression and $L_i(k, p_1, \cdots, p_J), i=1,
\cdots, I$ are linear
combinations of  $k$ and the  $p_j$'s.
   \\
  The definitions we adopt here are  inspired by work of Lesch and Pflaum
 \cite{LP} on traces of parametric pseudodifferential operators. Even though
 our symbols with affine parameters are not strongly parametric symbols as are
 the
 ones  used in their work, their general approach can be adapted to our
 context, thereby offering an interpretation in terms of iterated integrals of
 symbols with linear constraints of computations carried out by physicists to
 evaluate Feynman diagrams.  \\
 The affine parameters here play the role of external
   momenta in physics and we describe
   two ways of regularising 
    integrals of the type (\ref{eq:Feynmanint}). We discuss the Taylor
    truncation method implemented by physicists which gives rise to  
regularised integrals defined  modulo polynomials. \\
Following conventions used in the pseudodifferential litterature, we choose to
denote by $\eta_i$ the external parameters.
\begin{lem} 
 For   
any $\sigma_i\in CS^{a_i, *} (R^n) $ such that
 $\sum_{i=1}^k{\rm Re}( a_i)<- n$ where  
$a_i$ is the order  of $\sigma_i$, then 
 the  integral with affine parameters $\eta_i\in \R^n,i=1, \cdots, k$
$$ \int_{\R^n}\,  \left(\sigma_1\otimes
\cdots \otimes \sigma_k\right)(\xi+\eta_1, \cdots, \xi+\eta_k)\, d\,\xi$$   is well defined 
\end{lem}
{\bf Proof:} Since the $\sigma_i$'s are symbols we have
$\vert\sigma_i(\xi+\eta_i)\vert \leq C_i\, \langle
\xi+\eta_i\rangle^{a_i}$ for some $C_i\in \R_+$ and where we have set $\langle \zeta\rangle:= \sqrt{1+\vert \zeta\vert^2}$. 
But the integral 
$\int_{\R^{n}}\,\prod_{i=1}^k \langle \xi\, \dbar \xi+\eta_i\rangle^{a_i}$ is convergent
whenever $\sum_{i=1}^k{\rm Re} (a_i)<-n$ hence the result. \endsquare\\ \\
\begin{rk}  A typical example of such an integral is:
$$p\mapsto \int_{\R^n} \frac{P(k,
  p)}{(k^2+m^2)^{s_1}((k-p)^2+m^2)^{s_2}}\, dk$$
where $P$ is a polynomial expression and the $s_i$ are complex numbers with
real part  chosen large enough for
the integral to converge. Here, we have adopted the physicists notation $k^2$
for $\vert k\vert^2$. Writing  the polynomial $P(k,p)$ as a polynomial
$\sum_\alpha a_\alpha(p)k^\alpha$ in
$k$ with  coefficients depending polynomially on $p$, we can rewrite the integrand
 as a finite linear combination (with
$p$-dependent coefficients) of $k\mapsto \frac{
  k^\alpha}{(k+m^2)^{s_1}((k-p)^2+m^2)^{s_2}}$ each of  which reads
$k\mapsto (\tau_\alpha\otimes \sigma_1\otimes \sigma_2)(k, k, k-p)$
where we have set $\tau_\alpha(k):= k^\alpha$,
$\sigma_i(k)=\frac{1}{(k^2+m^2)^{s_i}}$. 
 \end{rk}
Let us now describe a  procedure to regularise integrals with
affine parameters  used
 by physicists to compute Feynman integrals. The idea is to truncate the Taylor series in the
$\eta_i$ (which correspond to external momenta in physics)  about the origin
at a high enough order. With the 
 notations of \cite{Sp}, we denote by ${\cal M}$ this truncation and define an
 alternative cut-off regularised integral (\ref{eq:cutoffextmomenta})
$$\altcutoffint_{\R^n} \left(\sigma_1\otimes
\cdots \otimes \sigma_k\right)(\xi+\eta_1, \cdots, \xi+\eta_k)\, d\,\xi=
\int_{\R^n} \left(I-{\cal M}\right)\left(\sigma_1\otimes
\cdots \otimes \sigma_k\right)(\xi+\eta_1, \cdots, \xi+\eta_k)\, d\,\xi.$$
 In physics the order of truncation (given here by $N$)
is chosen according to the superficial degree of divergence of the
diagram. In contrast, here we do not fix the order of truncation as it will
soon appear that it  can be chosen arbitrarily large. \\
It turns out that this regularised integral is defined ``up to polynomials''
in the components of the external parameters and that it coincides with the
previously defined cut-off regularised integral ``up to polynomials'' in these
external parameters.
\begin{prop}
For  any  
any $\sigma_i\in CS^{a_i, *} (R^n) $, 
 {\rm modulo polynomials in the components of the parameters
$\eta_1\in \R^n, \cdots, \eta_k\in \R^n$ } the expression:
\begin{eqnarray}\label{eq:cutoffextmomenta}
&{}&\altcutoffint_{\R^n} \left(\sigma_1\otimes
\cdots \otimes \sigma_k\right)(\xi+\eta_1, \cdots, \xi+\eta_k)\, d\,\xi\nonumber\\
&:=& \int_{\R^n} \, \left[\left(\sigma_1\otimes
\cdots \otimes \sigma_k\right)(\xi+\eta_1, \cdots, \xi+\eta_k)\right.\nonumber\\
&{}&\left. -\sum_{\vert \beta\vert=0}^N \partial_1^{\beta_1}\cdots \partial_k^{\beta_k}\left(\sigma_1\otimes
\cdots \otimes \sigma_k\right)(\xi, \cdots, \xi)\, \frac{\eta_1^{\beta_1}\cdots \eta_k^{\beta_k}}{\beta_1!\cdots \beta_k!}\right]\, d\,\xi
\end{eqnarray}
is well defined  and coincides  {\rm modulo polynomials in the components of the parameters
$\eta_1, \cdots, \eta_k$ } with the ordinary integral whenever
$\sum_{i=1}^k{\rm Re} (a_i)<- n$. Here $\beta_1, \cdots, \beta_k$ are
multiindices in $\N^n$ and we have set $\vert \beta\vert:= \sum_{i=1}^k\vert \beta_i\vert$. 
\end{prop}
 {\bf Proof:} A Taylor expansion of $\left(\sigma_1\otimes
\cdots \otimes \sigma_k\right)(\xi+\eta_1, \cdots, \xi+\eta_k)$ in $\eta= (\eta_1, \cdots, \eta_k)$ at $\eta=0$ yields
\begin{eqnarray*}
&{}& \left(\sigma_1\otimes
\cdots \otimes \sigma_k\right)(\xi+\eta_1, \cdots, \xi+\eta_k)\\
&=&\sum_{\vert \beta\vert=0}^N \partial_1^{\beta_1}\cdots \partial_k^{\beta_k}\left(\sigma_1\otimes
\cdots \otimes \sigma_k\right)(\xi, \cdots, \xi)\,\,
\frac{\eta_{1}^{\beta_{1}}\cdots\eta_{n}^{\beta_{k}}}{\beta_{1}!\cdots
    \beta_{k}!}\\
&+& R_N(\xi, \eta_1, \cdots, \eta_k).
\end{eqnarray*}
Since the real part of the total order $\sum_{i=1}^k a_i-\vert \beta\vert$ of
$\partial^\beta (\sigma_1\otimes \cdots \otimes\sigma_k)$  decreases as $\vert \beta\vert$ increases, 
the remainder term $R_N(\xi, \eta_1, \cdots, \eta_k)$ lies in $L^1(\R^n)$ provided $N$ is chosen large enough. 
The integral 
\begin{eqnarray*}
&{}&\int_{\R^n}\, \left[\left(\sigma_1\otimes
\cdots \otimes \sigma_k\right)(\xi+\eta_1, \cdots, \xi+\eta_k)\right.\\
&-&\left. \sum_{\vert \beta\vert=0}^N \partial_1^{\beta_1}\cdots \partial_k^{\beta_k}\left(\sigma_1\otimes
\cdots \otimes \sigma_k\right)(\xi, \cdots, \xi)\,\,
\frac{\eta_1^{\beta_1}\cdots \eta_k^{\beta_k}}{\beta_1!\cdots \beta_k!}\right] \, d\,\xi
\end{eqnarray*} therefore makes sense for large enough $N$. 
A modification of $N$ only modifies the expression by a polynomial in $\eta_1,
\cdots, \eta_k$ so that the expression is well-defined modulo polynomials. \\
When $\sum_{i=1}^n {\rm Re}(a_i)<-n$ the integral $\int_{\R^n}\left(\sigma_1\otimes
\cdots \otimes \sigma_k\right)(\xi+\eta_1, \cdots, \xi+\eta_k) \, d\,\xi$ converges by the above
lemma and hence so does $\xi \mapsto R_N(\xi, \eta_1, \cdots, \eta_k)$ lie in
$L^1(\R^n)$. The Taylor expansion then yields  (\ref{eq:cutoffextmomenta})
with the cut-off regularised integral on the l.h.s. replaced by an ordinary
integral. It follows that the cut-off regularised  integral $\altcutoffint$
coincides (modulo polynomials in the components of $\eta_i$) with the usual
integral whenever the
integrand converges. \endsquare\\\\
 The following result shows that derivations w.r. to the parameters
  commute with the regularised integral $\altcutoffint$.
 \begin{thm}\label{thm:parameterdeptintegrals}
Modulo polynomials in the components of the parameters
$\eta_1, \cdots, \eta_k$ we have 
\begin{eqnarray*}&{}&\partial_{\eta}^{\gamma}\altcutoffint_{\R^n} \left(\sigma_1\otimes
\cdots \otimes \sigma_k\right)(\xi+\eta_1, \cdots, \xi+\eta_k)\, d\,\xi\\
&=&
\altcutoffint_{\R^n}\partial_{ \eta}^{\gamma} \left(\sigma_1\otimes
\cdots \otimes \sigma_k\right)(\xi+\eta_1, \cdots, \xi+\eta_k)\, d\,\xi,\quad
\forall \gamma \in \N^k
\end{eqnarray*}
where for the multiindex $\gamma=(\gamma_1, \cdots, \gamma_k)$ we have set
$\partial_\eta^\gamma:= \partial_{\eta_1}^{\gamma_1}\cdots \partial_{\eta_k}^{\gamma_k}$.
Provided $\vert \gamma\vert=\gamma_1+ \cdots +\gamma_k$ is chosen large enough so
 that  the integrand $\partial_{\eta_1}^{\gamma_1}\cdots \partial_{\eta_k}^{\gamma_k} \left(\sigma_1\otimes
\cdots \otimes \sigma_k\right)(\xi+\eta_1, \cdots, \xi+\eta_k)$ lies in
$L^1(\R^n)$  we have:
\begin{eqnarray} \label{eq:differentiatedaltcutoff}
&{}&\partial_{\eta}^{\gamma}\altcutoffint_{\R^n} \left(\sigma_1\otimes
\cdots \otimes \sigma_k\right)(\xi+\eta_1, \cdots, \xi+\eta_k)\, d\,\xi\\
&=& \int_{\R^n} \partial_{\eta}^{\gamma}\left(\sigma_1\otimes
\cdots \otimes \sigma_k\right)(\xi+\eta_1, \cdots, \xi+\eta_k)\, d\,\xi \quad
{\rm mod} \quad {\rm polynomials}\nonumber.
\end{eqnarray}
\end{thm}
{\bf Proof:} We prove the result for $\gamma=(0, \cdots, 0, \gamma_i, 0,
\cdots 0)$ from which the general statement then easily follows.  Whenever $\sum_{i=1}^k{\rm Re}(a_i)<-n$ 
we have  $$\partial_{\eta_i}^{\gamma_i}\int_{\R^n} \left(\sigma_1\otimes
\cdots \otimes \sigma_k\right)(\xi+\eta_1, \cdots, \xi+\eta_k)\, d\,\xi=\int_{\R^n} \partial_{\eta_i}^{\gamma_i}\left(\sigma_1\otimes
\cdots \otimes \sigma_k\right)(\xi+\eta_1, \cdots, \xi+\eta_k)\, d\,\xi.$$
Hence by (\ref{eq:cutoffextmomenta}),  modulo polynomials in the $\eta_i$ we have
\begin{eqnarray*}
&{}&\partial_{\eta_i}^{\gamma_i}\altcutoffint_{\R^n} \left(\sigma_1\otimes
\cdots \otimes \sigma_k\right)(\xi+\eta_1, \cdots, \xi+\eta_k)\, d\,\xi\\
&=&\partial_{\eta_i}^{\gamma_i} \left[ \int_{\R^n} \,\left(\sigma_1\otimes
\cdots \otimes \sigma_k\right)(\xi+\eta_1, \cdots, \xi+\eta_k)\right.\\
&&\left.-\sum_{\vert \beta\vert=0}^N \partial_1^{\beta_1}\cdots \partial_k^{\beta_k}\left(\sigma_1\otimes
\cdots \otimes \sigma_k\right)(\xi, \cdots, \xi)\,\,
\frac{\eta_1^{\beta_1}\cdots \eta_k^{\beta_k}}{\beta_1!\cdots
  \beta_k!}\right]\, d\,\xi\\
\end{eqnarray*}
\begin{eqnarray*}
&=& \int_{\R^n} \,\partial_{\eta_i}^{\gamma_i} \left[\left(\sigma_1\otimes
\cdots \otimes \sigma_k\right)(\xi+\eta_1, \cdots, \xi+\eta_k)\right.\\
&&\left.-\sum_{\vert \beta\vert=0}^N \partial_1^{\beta_1}\cdots \partial_k^{\beta_k}\left(\sigma_1\otimes
\cdots \otimes \sigma_k\right)(\xi, \cdots, \xi)\, \frac{\eta_1^{\beta_1}\cdots \eta_k^{\beta_k}}{\beta_1!\cdots \beta_k!}\right]\, d\,\xi\\
&=& \int_{\R^n} \,\left[\partial_{\eta_i}^{\gamma_i} \left(\sigma_1\otimes
\cdots \otimes \sigma_k\right)(\xi+\eta_1, \cdots, \xi+\eta_k)\right.
\\
&&\left. -\sum_{\vert \beta\vert=0}^N \partial_1^{\beta_1}\cdots \partial_k^{\beta_k}\left(\sigma_1\otimes
\cdots \otimes \sigma_k\right)(\xi, \cdots, \xi)\, \partial_{\eta_i}^{\gamma_i}  \frac{\eta_1^{\beta_1}\cdots \eta_k^{\beta_k}}{\beta_1!\cdots \beta_k!}\right]\, d\,\xi\\
&=&\altcutoffint_{\R^n} \partial_{\eta_i}^{\gamma_i}\left(\sigma_1\otimes
\cdots \otimes \sigma_k\right)(\xi+\eta_1, \cdots, \xi+\eta_k)\, d\,\xi \quad
{\rm mod} \quad {\rm polynomials} .
\end{eqnarray*}
This proves the first part of the statement.\\
Since differentiation w.r. to the $\eta_i$ decreases the total order of the symbol, for large enough $\vert \gamma\vert$, the integrand $\partial_{\eta_1}^{\gamma_1}\cdots \partial_{\eta_k}^{\gamma_k} \left(\sigma_1\otimes
\cdots \otimes \sigma_k\right)(\xi+\eta_1, \cdots, \xi+\eta_k)$ lies in
$L^1(\R^n)$ and  we can write:
\begin{eqnarray*}
&{}&\partial_{\eta_1}^{\gamma_1}\cdots \partial_{\eta_k}^{\gamma_k}\altcutoffint_{\R^n} \left(\sigma_1\otimes
\cdots \otimes \sigma_k\right)(\xi+\eta_1, \cdots, \xi+\eta_k)\, d\,\xi\\
&=& \int_{\R^n} \partial_{\eta_1}^{\gamma_1}\cdots \partial_{\eta_k}^{\gamma_k}\left(\sigma_1\otimes
\cdots \otimes \sigma_k\right)(\xi+\eta_1, \cdots, \xi+\eta_k)\, d\,\xi \quad
{\rm mod} \quad {\rm polynomials}
\end{eqnarray*}
which ends the proof of the proposition.
\endsquare\\ \\
 In certain situations the maps:
$$\eta_i\mapsto \altcutoffint_{\R^n} \left(\sigma_1\otimes
\cdots \otimes \sigma_k\right)(\xi+\eta_1, \cdots, \xi+\eta_k)\, d\,\xi$$ to
define symbols (modulo polynomials in $\eta_i$), in which case one cannot
expect the latter to be classical but 
rather log-polyhomogeneous. In that case, one can further integrate in the
 parameter $\eta_i$  using  cut-off integration.\\ 
The ambiguity that
arises from having expressions defined  ``modulo polynomials'' in the external
parameters disapears after
cut-off integration in these parameters  as a
result  of the fact that the cut-off regularised integral
$\cutoffint_{\R^n}$ vanishes on  polynomials.\\
Consequently, the order of truncation at which the Taylor expansion was
originally taken in the external parameters does not matter as long as it is
chosen large enough: extra terms in the Taylor expansion  are polynomials in
the external parameters and hence vanish after cut-off integration in these parameters.
\begin{cor}\label{cor:parameterdeptintegrals}
 Whenever $\eta_i\mapsto \altcutoffint_{\R^n} \left(\sigma_1\otimes
\cdots \otimes \sigma_k\right)(\xi+\eta_1, \cdots, \xi+\eta_k)\, d\,\xi$ lies in 
$CS^{*,*} (\R^n)$ (modulo polynomials in $\eta_i$),  the double cut-off regularised integral: $$\cutoffint_{\R^n} \left(\altcutoffint_{\R^n} \left(\sigma_1\otimes
\cdots \otimes \sigma_k\right)(\xi+\eta_1, \cdots, \xi+\eta_k)\,
d\,\xi\right)\,  d\eta_i $$
is well defined modulo polynomials in the remaining $\eta_j,j\neq i$.\\
 If $\eta_i\mapsto \altcutoffint_{\R^n} \left(\sigma_1\otimes
\cdots \otimes \sigma_k\right)(\xi+\eta_1, \cdots, \xi+\eta_k)\, d\,\xi$ moreover has
non integer order, then for large enough $\vert \gamma_i\vert$ we have 
\begin{eqnarray*}
&{}&\cutoffint_{\R^n}  \left(\altcutoffint_{\R^n} \left(\sigma_1\otimes
\cdots \otimes \sigma_k\right)(\xi+\eta_1, \cdots, \xi+\eta_k)\,
d\,\xi\right)\, d\eta_i \\
&=& (-1)^{\vert\gamma_i\vert}\int_{\R^n}\frac{\eta_i^{\gamma_i}}{\gamma_i!}  \left(\int_{\R^n}\partial_{\eta_i}^{\gamma_i} \left(\sigma_1\otimes
\cdots \otimes \sigma_k\right)(\xi+\eta_1, \cdots, \xi+\eta_k)\,
d\,\xi\right) \, d\eta_i \\ 
\end{eqnarray*}
where the cut-off integral $\cutoffint_{\R^n}$ has now been replaced by an
ordinary integral.
\end{cor}
{\bf Proof:} The first part of the statement follows from the fact that the
cut-off regularised integral $\cutoffint_{\R^n}$ vanishes on polynomials (see
Proposition  \ref{prop:polynomials}). The
second part of the statement follows from integration by parts property (see
Proposition \ref{prop:Stokes}) for
the cut-off integral on non integer order symbols combined with
(\ref{eq:differentiatedaltcutoff}). 
Indeed, we have: 
\begin{eqnarray*}
&{}&\cutoffint_{\R^n}  \left(\altcutoffint_{\R^n} \left(\sigma_1\otimes
\cdots \otimes \sigma_k\right)(\xi+\eta_1, \cdots, \xi+\eta_k)\,
d\,\xi\right)\, d\eta_i \\
&=& (-1)^{\vert \gamma_i\vert} \cutoffint_{\R^n}  \partial_{\eta_i}^{\gamma_i}\ \left(\altcutoffint_{\R^n} \left(\sigma_1\otimes
\cdots \otimes \sigma_k\right)(\xi+\eta_1, \cdots, \xi+\eta_k)\,
d\,\xi\right)\, d\eta_i \quad  {\rm (by} \quad {\rm 
Proposition }\quad {\rm \ref{prop:Stokes})} \\
&=& (-1)^{\vert \gamma_i\vert} \, \cutoffint_{\R^n}  \left(\int_{\R^n}\partial_{\eta_i}^{\gamma_i} \left(\sigma_1\otimes
\cdots \otimes \sigma_k\right)(\xi+\eta_1, \cdots, \xi+\eta_k)\,
d\,\xi\right)\, d\eta_i \quad ({\rm by}\quad \quad {\rm (\ref{eq:differentiatedaltcutoff})}). \\
\end{eqnarray*}\endsquare\vfill \eject \noindent
\section*{Part 2:  Renormalised multiple integrals of  
  symbols with linear constraints}
The aim of this second part of the paper is  to define renormalised multiple integrals with
linear constraints. Instead of
iterating regularised integrals as one might do for convergent integrals, we
renormalise multiple  integrals as a whole in the spirit of  Connes and
Kreimer's approach to renormalisation of Feynman diagrams, keeping in mind
that to such a diagram corresponds a multiple integral with affine constraints. \\ 
It is useful to first recall how multiple integrals of symbols  without constraints can
be renormalised using a Birkhoff factorisation.
\section{ Integrals of tensor products of symbols revisited}
We report on and extend results of   \cite{MP} concerning  integrals of tensor products
of symbols, i.e. multiple integrals without constraints. \\ 
  Following \cite{MP}, let us consider
the tensor algebra  of log-polyhomogeneous symbols:
$${\cal T}\left(CS (\R^n)\right):= \bigoplus_{k=0}^\infty
\hat \otimes^kCS (\R^n)$$
built on the algebra $CS (\R^n)$ of log-polyhomogeneous  symbols  on
$\R^n$. Here $\hat{}$ denotes the Grothendieck completion. \\
The cut-off regularised integral being  continuous on the subspace $CS
^{a}(\R^n)$ of classical  symbols  on
$\R^n$ with constant order $a$ for any fixed $a\in \C$, it can be extended
by continuity and (multi-) linearity 
to the tensor algebra ${\cal T}\left(CS (\R^n)\right)$. 
\begin{defn} \cite{MP} The cut-off  regularised integral $\cutoffint_{\R^n}$
   defined on $CS(\R^n)$  extends to a character: 
\begin{eqnarray*} {\cal T}\left(CS(\R^n)\right)&\to & \C\\
\sigma_1\otimes \cdots \otimes \sigma_k&\mapsto& \cutoffint_{\R^{nk}}\sigma_1\otimes
\cdots\otimes \sigma_k:=\prod_{i=1}^k  \cutoffint_{\R^n} \sigma_i.
\end{eqnarray*}
\end{defn}
As an immediate consequence of these definitions and the previous results on
ordinary cut-off regularised integrals we have the following meromorphicity
result (This is a slight generalisation of results in \cite{MP}).
\begin{lem}\label{lem:holcutofftensor}
 Given a continuous holomorphic regularisation procedure ${\cal R}$ on
 $CS (\R^n)$, for  any  $ \sigma_i \in CS (\R^n), i=1,
 \cdots, k$ 
the map $\underline z\mapsto \cutoffint_{R^{nk}}{\cal R}(\sigma_1)(z_1)\otimes
\cdots\otimes{\cal R}( \sigma_k)(z_k)$
is meromorphic  with simple poles 
 and  we have the following factorisation property as an equality of
 meromorphic functions: 
$$  \cutoffint_{\R^{nk}}{\cal R}(\sigma_1)(z_1)\otimes
\cdots\otimes {\cal R}(\sigma_k)(z_k)
= \prod_{i=1}^k\cutoffint_{\R^n}{\cal R}( \sigma_i)(z_i).$$ 
\end{lem}
 \begin{defn} Given a continuous holomorphic regularisation ${\cal R}:\sigma
   \mapsto \sigma(z)$  the regularised integral $\cutoffint^{\cal R}$
   defined on $CS(\R^n)$ extends to a character: 
\begin{eqnarray*} {\cal T}\left(CS(\R^n)\right)&\to & \C\\
\sigma_1\otimes \cdots \otimes \sigma_k&\mapsto& \cutoffint_{\R^{nk}}^{\cal R}\sigma_1\otimes
\cdots\otimes \sigma_k:=\prod_{i=1}^k  \cutoffint^{\cal R} \sigma_i.
\end{eqnarray*}
It  coincides with the ordinary integral when the
integrands $\sigma_i$ all lie  in $L^1(\R^{n})$:
$$\cutoffint^{{\cal R}}_{\R^{nk}}\sigma_1\otimes
\cdots\otimes \sigma_k= \int_{\R^{nk}}\sigma_1\otimes
\cdots\otimes \sigma_k.$$

\end{defn}
\begin{rk}Unless the partial sums of the orders of the symbols $\sigma_i$ are non
integer valued or the integral converges, one is to expect that 
$$\cutoffint_{\R^{nk}}^{\cal R}\sigma_1\otimes
\cdots\otimes \sigma_k\neq   {\rm fp}_{z=0}\cutoffint_{\R^{nk} }
{\cal R}\left(\sigma_1\right)(z)\otimes
\cdots\otimes {\cal R}\left( \sigma_k\right)(z)$$
since the finite part of a product of meromorphic functions, namely of  the maps
$z\mapsto \int_{\R^n} {\cal R}\left( \sigma_i\right)(z)$ with  $i\in \{1,
\cdots, I\}$,   does not generally
coincide with the product of the finite parts of these functions. 
\end{rk}
However, if one insists on setting $z_i=z$ for $i\in \{1, \cdots,
  I\}$ then one can implement a
renormalisation procedure using Birkhoff factorisation to take care of the problem mentioned in the above
remark\footnote{Such a situation  arises in physics when using dimensional
  regularisation for the parameter $z$  used to complexify the dimension
  thereby modifies the integrands via a common complex parameter $z$.}. \\
For this purpose we  equip the  tensor algebra ${\cal T}\left(CS (\R^n)\right):=
\bigoplus_{k=0}^\infty {\cal T}^k\left(CS (\R^n)\right)$ where we
have set 
${\cal T}^k\left(CS (\R^n)\right):= \hat \otimes^k CS_{\rm c.c.}(\R^n)$
 with the ordinary tensor product
$\otimes$  and the 
 deconcatanation coproduct:
\begin{eqnarray*}\Delta:{\cal T}\left(CS (\R^n)\right) &\to&
  \bigoplus_{p+q=L}\left({\cal T}^p\left(CS (\R^n)\right)\bigotimes{\cal T}^q\left(CS (\R^n)\right)\right) \\
\sigma_1\otimes \cdots\otimes \sigma_k&\mapsto & \sum_{\{i_1, \cdots,i_{l^\prime}\}\subset \{1, \cdots, k\}} 
\left(\sigma_{i_1}\otimes \cdots\otimes \sigma_{i_{k^\prime}}\right)\bigotimes
 \left(\sigma_{i_{l+1}}\otimes \cdots\otimes \sigma_{i_k}\right)\\
\end{eqnarray*}
where $\{ i_{k^\prime+1}, \cdots, i_k\}$ is the complement in 
$\{1, \cdots, k\}$ of the set $\{i_1, \cdots,i_{k^\prime}\}$.\\
Let us recall the following well-known results (see e.g. \cite{M}):
\begin{lem}  ${\cal H}^0:= \left({\cal T}\left(CS ^{*, *}(\R^n)\right), \otimes, \Delta\right)$
  is a graded (by the natural grading on tensor products) cocommutative connected  Hopf algebra.
\end{lem}
\begin{rk} \cite{M} This corresponds to  the natural structure of cocommutative Hopf algebra on the
  tensor algebra of any vector space $V$ with the coproduct $ \Delta$  given
  by the unique
  algebra morphism from  ${\cal T}(V)\to {\cal T}(V)\otimes {\cal T}(V)$ such
  that $\Delta(1)= 1\otimes 1$ and $ \Delta(x)= x\otimes 1+ 1\otimes x$. 
\end{rk}
{\bf Proof:} We use Sweedler's notations and write in a compact form
$$ \Delta\sigma= \sum_{(\sigma)} \sigma_{(1)}\otimes \sigma_{(2)}.$$
\begin{itemize}
\item
The coproduct $\Delta$ is clearly compatible with the filtration.
\item The coproduct  $\Delta$ is  cocommutative for we have 
$\tau_{12} \circ \Delta= \Delta$ where $\tau_{ij}$ is the flip on the
$i$-th and $j$-th entries:
\begin{eqnarray*}
\tau_{12} \circ \Delta (\sigma)
&=& \tau_{12}\left( \sum_{({\sigma})}\sigma_{(1)}\otimes \sigma_{(2)}\right)\\
&=& \sum_{({\sigma})}\sigma_{(2)}\otimes \sigma_{(1)}\\
&=&  \Delta (\sigma).
 \end{eqnarray*}
\item The coproduct  $\Delta$ is   coassociative since 
\begin{eqnarray*}(\Delta\otimes 1)\circ \Delta (\sigma)
&=& \sum_{({\sigma})}\left( \sigma_{(1:1)}\otimes
\sigma_{(1:2)}\right)\otimes \sigma_{(2)}\\
&=&  \sum_{(\sigma)} \sigma_{(1)}\otimes \left(\sigma_{(2:1)}
  \otimes\sigma_{(2:2)}\right)\\
&=& (1\otimes \Delta)\circ \Delta (\sigma).
\end{eqnarray*}
\item  The co-unit $\varepsilon$ defined by $\varepsilon(1)=1$ is an algebra
morphism.
\item The coproduct  $\Delta$ is compatible with  the product
$\otimes$. 
\begin{eqnarray*}
 \Delta  \circ m \left(\sigma\otimes\sigma^\prime \right)
&=& \sum_{(\sigma\otimes
  \sigma^\prime)}(\sigma\otimes\sigma^\prime)_{(1)}\bigotimes(\sigma\otimes\sigma^\prime)_{(2)} \\
&=&( m\bigotimes m )\circ  \tau_{23}\circ \left[(\sigma_{(1)}\otimes \sigma_{(2)})
 \bigotimes (\sigma^\prime_{(1)}\otimes \sigma^\prime_{(2)}) \right]\\
&=& ( m\otimes m )\circ  \tau_{23}\circ(\Delta \otimes \Delta) \left(\sigma \otimes
  \sigma^\prime\right).\\
\end{eqnarray*}
\end{itemize}
\endsquare \\ \\
 We derive the following meromorphicity result as an easy consequence of   Lemma \ref{lem:holcutofftensor}: 
\begin{prop}\label{prop:holcutofftensor}
Given a continuous holomorphic regularisation procedure ${\cal R}$ on ${\cal H}^0$, 
the map 
\begin{eqnarray*}
\Phi^{\cal R}:\left({\cal H}^0 ,\otimes\right)& \to &{\cal M}(\C)\\
\sigma_1\otimes\cdots\otimes\sigma_k&\mapsto &  \cutoffint_{\R^{nk}}{\cal R}(\sigma_1)\otimes
\cdots\otimes{\cal R}( \sigma_k) ,\\
\end{eqnarray*}
where ${\cal M}(\C)$ denotes the algebra of meromorphic functions is well
defined and induces  
 an algebra  morphism  on  $\left({\cal H}^0, \otimes\right)$.
\end{prop}
A Birkhoff factorisation procedure then yields a complex valued character. 
\begin{thm}\label{thm:holcutofftensor}
A continuous holomorphic regularisation procedure ${\cal R}$ on $CS(\R^n)$ gives rise to a character  on the Hopf algebra  $\left({\cal H}^0
, \otimes\right):$
\begin{eqnarray*}
\phi^{\cal R}:{\cal H}^0& \to &\C\\
\sigma_1\otimes \cdots\otimes\sigma_k&\mapsto &  \cutoffint^{{\cal R},{\rm
ren}}_{\R^{nk} }\sigma_1\otimes
\cdots\otimes \sigma_k\\
\end{eqnarray*} which therefore   coincides with the extended regularised
integral  $\cutoffint^{{\cal R}}$  on ${\cal T}(CS(\R^n))$.\\
 In particular we
have the following multiplicative property: 
\begin{eqnarray*}
&{}&\cutoffint^{{\cal R},{\rm ren}}_{\R^{nk} }\left(\sigma_1\otimes
\cdots\otimes \sigma_k\right)\otimes \left(\sigma^\prime_1\otimes
\cdots\otimes \sigma^\prime_{k^\prime}\right)\\
&=& \left(\cutoffint^{{\cal R},{\rm ren}}_{\R^{nk}}\sigma_1\otimes
\cdots\otimes\sigma_k\right)\cdot \left(\cutoffint^{{\cal R},{\rm ren}}_{\R^n\times \cdots \times \R^n }\sigma^\prime_1\otimes
\cdots\otimes \sigma^\prime_{k^\prime}\right).
\end{eqnarray*}
\end{thm}
{\bf Proof:}
Birkhoff factorisation  combined with a  minimal substraction scheme
 yields the existence of a character  on the connected
 filtered commutative Hopf  algebra  ${\cal H}^0 $
 \cite{M} (Theorem II.5.1) 
$$ \Phi_+^{\cal R}: \left({\cal H}^0 , \otimes\right) \longrightarrow  {\rm Hol}(\C)$$corresponding to  the holomorphic part in the unique Birkhoff decomposition  $\Phi^{{\cal R}}= \left(\Phi_-^{{\cal R}}\right)^{*-1}
 \star \Phi^{{\cal R}}_+$  of
 $\Phi^{{\cal R}}$, $\star$ being the convolution product on the Hopf
 algebra.  Here ${\rm  Hol}(\C)$ is the
 algebra of holomorphic functions.
Its value $\phi^{\cal R}:= \Phi_+^{\cal R}(0)$ at $z=0$ yields in
turn a  character  $\phi^{\cal R}:\left( {\cal H}^0   , \otimes\right)
\to  \C$
$$\phi^{\cal R}\left(\sigma_1\otimes \cdots \otimes\sigma_k\right)=
\cutoffint^{{\cal R},{\rm ren}}_{\R^{nk}} \sigma_1\otimes \cdots \otimes \sigma_k$$
 which extends   the map given by the
ordinary iterated integral. The multiplicativity of these renormalised
integrals $\cutoffint^{{\cal R},{\rm ren}}_{\R^{nk}}$ w.r. to tensor products follows from the character
 property of $\phi^{\cal R}$. Since $\cutoffint_{\R^n}^{\cal R}$ extends in a
 unique way to a character on ${\cal T}(CS(\R^n))$, the character $
 \cutoffint^{{\cal R},{\rm ren}}_{\R^{nk}} $ coincides with the afore
 defined extension  $\cutoffint_{\R^{nk}}^{\cal R}$.  \endsquare
\section{ Linear constraints in terms of matrices}
Adding in  linear constraints is carried out introducing
  matrices. To a matrix $B$ with real coefficients  $$B= \left( \begin{array}{llcl} b_{11}&b_{12}& \cdots&b_{1L}\\
                                    \cdots&\cdots &\cdots &\cdots\\
				    b_{I1}& b_{I2} &\cdots & b_{IL} \end{array}\right) $$
and  symbols $\sigma_i \in CS(\R^n), i=1, \cdots, I$ we associate 
the  map   $$(\xi_1, \cdots, \xi_L)\mapsto (\sigma_1\otimes \cdots \otimes \sigma_I)\circ B(\xi_1, \cdots, \xi_L):=\sigma_1  \left(\sum_{l=1}^L b_{1l}\xi_l\right) \cdots
\sigma_I\left(\sum_{l=1}^Lb_{Il}\xi_l\right)$$ and we want to investigate the
corresponding multiple integral with linear constraints:
$$\int_{\R^{nL}} (\sigma_1\otimes \cdots \otimes \sigma_I)\circ B(\xi_1,
\cdots, \xi_L)\, d\xi_1\cdots d\xi_L.$$
\begin{rk}\begin{enumerate}
\item A permutation $\tau \in \Sigma_I$ on the lines of $B$ amounts to 
  relabelling  the symbols $\sigma_i$ in the tensor product.
\item A permutation $\tau\in \Sigma_L$ on the columns of $B$ amounts to 
  relabelling  the variables $\xi_l$. 
\end{enumerate}
\end{rk}
 \begin{ex} \label{ex:Feynmanmatrix}Take $I=3, L=2$ and 
   $\sigma_i(\xi)= \frac{1}{m^2+\vert\xi\vert^2} \quad \forall i=1, 2, 3$. Then 
 $$ \frac{1}{m^2+\vert \xi_1\vert^2} \  \frac{1}{m^2+\vert \xi_1+ \xi_2\vert^2}\, 
\frac{1}{m^2+\vert \xi_2\vert^2}\\
=\left((\sigma_1\otimes\sigma_2\otimes \sigma_3)\circ B \right)(\xi_1, \xi_2,\xi_3)$$
where 
$B= \left( \begin{array}{cc} 1&0\\
                                    1&1\\
				    0&1\end{array}\right).$
\end{ex}
 Feynman diagrams give rise to integrals with integrands of this type up to the
  fact that here we omit external momenta; allowing for external momenta would
  lead to affine constraints,  a case which lies out of the scope of
  this article but which we hope to investigate in forthcoming work. Constraints on the momenta follow from
  the conservation of momentum as it flows through the diagram and  
$L$ corresponds to the
  number of loops in the diagram. \\ \\
Given a holomorphic regularisation ${\cal R}:
\sigma\mapsto \sigma(z)$, we  extend it to $\tilde \sigma \circ B$ with $\tilde \sigma \in {\cal
  T}\left(CS (\R^n)\right)$ and $B$ a matrix by:
 $$\widetilde{{\cal R}}(\tilde \sigma)(\underline z) \circ B:= \left({\cal
   R}(\sigma_1)(z_1)\otimes \cdots\otimes  {\cal
   R}(\sigma_k)(z_k)\right)\circ B\quad \forall \underline z=(z_1, \cdots, z_k)\in \C^k$$
which we also write $\tilde \sigma(z)\circ B$ for short.
\begin{prop}\label{prop:holmultintsymb}Let  $\sigma_i\in CS(\R^n)$ of order $a_i$. Let ${\cal R}$ be a continuous holomorphic regularisation and let for $i=1, \cdots, I$,  $\alpha_i(z)$ denote the order of $\sigma_i(z)$
which we assume is affine $\alpha_i(z)= \alpha_i^\prime(0)z+ a_i$
with real coefficients and such that  $\alpha_i^\prime(0)<0$. \\
If a matrix  $B=(b_{il})$ of size $I\times L$ and rank $L$, the map
$$ \underline z\mapsto \int_{\R^{n^L}}\widetilde{{\cal
     R}}\left(\tilde \sigma\right) \circ B(\underline z)$$
is holomorphic  on  the domain  $D=\{\underline z\in \C^I,\quad {\rm
   Re}(z_i) > -\frac{a_i+n}{\alpha_i^\prime(0)}, \quad \forall i\in \{1,
 \cdots, I\} \}$.
 
\end{prop}
{\bf Proof:}  The symbol property of each $\sigma_i$ yields the existence of a
constant $C$ such that 
\begin{eqnarray*}
\vert \tilde \sigma(\underline z)\circ B (\xi_1, \cdots, \xi_L)\vert&\leq & C\,
\prod_{i=1}^I \langle\sum_{l=1}^L b_{il}\xi_l\rangle^{{\rm Re}(\alpha_i(z_i))}\\
&\leq & C  \,
\prod_{i=1}^I \langle\sum_{l=1}^L b_{il}\xi_l\rangle^{\alpha^\prime_i(0){\rm
    Re}(z_i)+a_i }\\
\end{eqnarray*}
where we have set $\langle \eta\rangle := \sqrt{1+\vert \eta\vert^2}$. \\
We  infer that for  ${\rm Re}(z_i)\geq \beta_i>0 $  
$$\vert\tilde  \sigma(\underline z)\circ B(\xi_1, \cdots, \xi_L)\vert\leq \prod_{i=1}^I
\langle\sum_{l=1}^L b_{il}\xi_l\rangle^{\alpha^\prime_i(0)\beta_i +a_i
}.$$
We claim that the map  $(\xi_1, \cdots, \xi_L)\mapsto \langle\sum_{l=1}^L b_{il}\xi_l\rangle^{\alpha^\prime_i(0)\beta_i +a_i
}$ lies in $L^1\left(R^{nL}\right)$ if $\beta_i>  -\frac{a_i+n}{\alpha_i^\prime(0)}$. Indeed, the matrix $B$
 being of rank $L$ by assumption, we can extract an invertible $L\times L$ matrix
$D$. Assuming for simplicity (and without loss of generality, since this
assumption  holds
up to permutation of the lines and columns) that it
corresponds to the $L$ first lines of $B$ we write: 
\begin{eqnarray*}
 \prod_{i=1}^I
\langle\sum_{l=1}^L b_{il}\xi_l\rangle^{\alpha^\prime_i(0)\beta_i +a_i}
&= &\prod_{i=1}^I \rho_i\circ B(\xi_1, \cdots, \xi_L)\\
&\leq& \prod_{i=1}^L \rho_i\circ D(\xi_1, \cdots, \xi_L)\\
\end{eqnarray*}
where we have set $\rho_i (\eta):=\langle
\eta\rangle^{\alpha^\prime_i(0)\beta_i +a_i} $ and used the fact that
$\rho_i(\eta)\geq 1$ and $\alpha^\prime_i(0)\beta_i +a_i<-n$. \\
But 
$$\int_{\R^{nL}} \otimes_{i=1}^L 
\rho_i\circ D= \vert {\rm det}D\vert 
 \prod_{i=1}^L\int_{\R^n} 
\rho_i$$
converges as a product of integrals of symbols of order $<-n$ so that   by
dominated convergence,
$\widetilde {\cal R}(\tilde \sigma)(\underline z)\circ B $ lies in $L^1(\R^{nL})$ for any
complex number $\underline z\in D$. \\
On the other hand,  the
derivative in $z$ of holomorphic symbols have same order as the original
symbols (see e.g. \cite{PS}), the differentiation possibly introducing logarithmic terms. Replacing
$\sigma_1(z_1), \cdots, \sigma_I(z_I)$ by $\partial^{\gamma_1}_{z_1}\sigma_1(z_1), \cdots,
 \partial^{\gamma_I}_{z_I}\sigma_I(z_I)$ in the above inequalities, we can infer by a similar procedure that for ${\rm Re}(z_i)\geq \beta_i> -\frac{a_i+n}{\alpha_i^\prime(0)} $
 the map 
$\underline z\mapsto \widetilde{\cal R}(\tilde \sigma)(\underline z)\circ B$ is uniformly bounded by an $L^1$ function.
The holomorphicity of $\underline z\mapsto \int_{\R^{nL}}
 \widetilde{\cal R}(\tilde \sigma)(\underline z)\circ B$ then
follows. \endsquare\\ \\
As a straightforward consequence, we infer the existence of a meromorphic extension of 
 the map   $z\mapsto \int_{\R^{nL}} \widetilde{\cal R}(\tilde
 \sigma) (\underline z)\circ B$ to the whole
 plane when $L=I$. 
\begin{cor}\label{cor:hol}Let  $\sigma_i\in CS(\R^n)$ be of order $a_i$. Let ${\cal R}$ be a continuous holomorphic
  regularisation which sends $\sigma_i$ to $\sigma_i(z)$ of order  $\alpha_i(z)= \alpha_i^\prime(0)z+ a_i$
with real coefficients and such that  $\alpha_i^\prime(0)<0$.
\\ Given an invertible matrix $B$ with $L$ columns,  the map
$$\underline z\mapsto \cutoffint_{\R^{nL}} \widetilde{\cal R}(\tilde\sigma)(\underline z)\circ B:=\vert
{\rm det}B^{-1}\vert\,\cutoffint_{\R^{nL}} \widetilde{\cal R}(\tilde\sigma)(\underline z) $$
yields a meromorphic extension to the whole complex plane  of the holomorphic
map $\underline z\mapsto
\int_{\R^{nL}}\widetilde{\cal R}(\tilde \sigma)(\underline z)\circ B$ defined 
 on  the domain  $D=\{\underline z\in \C^L, {\rm
   Re}(z_l) > -\frac{a_l+n}{\alpha_l^\prime(0)}, \quad\forall l\in \{1,
 \cdots, L\} \}$.
\end{cor}
{\bf Proof:} Let us set $\tilde \sigma(\underline  z):=\widetilde{\cal
  R}(\tilde \sigma(\underline z))$ as before. 
We know from the previous proposition that $\underline z\mapsto
\int_{\R^{nL}}\tilde \sigma(\underline z)\circ B$ defines 
 a holomorphic map on  
$D=\{ \underline z\in \C,{\rm Re}(z_i) > -\frac{a_i+n}{\alpha_i^\prime(0)},
\quad\forall i\in \{1, \cdots, I\}\}$. By a change of variable it
follows that in that region of the plane
$$ \int_{\R^{nL}}\tilde \sigma(z)\circ B:=\vert
{\rm det}B^{-1}\vert\,\int_{\R^{nL}}\tilde\sigma(\underline z). $$
But by the results of the previous sections we know that
$\underline z\mapsto \int_{\left(\R^n\right)^I}\tilde \sigma(\underline z)$
 extends to a meromorphic map on the whole complex plane given
by a cut-off regularised integral of a tensor product
of symbols $\underline z\mapsto \cutoffint_{\R^{nL}}\tilde \sigma(\underline z)$. Hence $$\underline z\mapsto 
 \cutoffint_{\R^{nL}}\tilde \sigma(\underline z)\circ B:=\vert {\rm
   det}B^{-1}\vert\,\int_{\R^{nL}}\tilde \sigma(\underline z)$$ provides a
meromorphic extension of the l.h.s. \endsquare
\section{ Multiple integrals of holomorphic families with
    linear constraints}
Let us   now show the existence of meromorphic extensions for integrals
$\underline z \mapsto \int_{\R^{nL}}
\widetilde{\cal R}(\tilde \sigma)(\underline z)\circ B$  built from more general matrices
$B$, where as before ${\cal R}$ is a continuous holomorphic regularisation and
$\tilde \sigma:=\sigma_1\otimes \cdots \otimes\tilde \sigma_I\in {\cal
  T}(CS(\R^n))$. \\ \\
The aim of this section is to prove the following result. \
\begin{thm} \label{thm:meromultintsymb} Let ${\cal R}: \sigma\mapsto \sigma(z)$ be a holomorphic
  regularisation procedure on $CS(\R_+)$ and let 
$\xi\mapsto \sigma_i(\xi):= \tau_i(\vert
\xi\vert)\in CS(\R^n), i=1, \cdots, I$ be  {\rm radial} polyhomogeneous symbols    of order $a_i$ which are
sent via ${\cal R}$ to  $\xi\mapsto \sigma_i(z)(\xi):= {\cal R}(\tau)(z)(\vert \xi\vert)$ of non constant affine order
$\alpha_i(z)=-qz_i+a_i$, for some positive real number $q$. For any  matrix $B$ of size $I\times L$ and rank $L$,   the map 
$$\underline z\mapsto
\int_{\R^{nL}}\widetilde {\cal R}(\tilde \sigma)(\underline z)\circ B$$
which is well defined and  holomorphic on the domain  $D=\{\underline z\in \C^I,\quad {\rm
   Re}(z_i) > -\frac{a_i+n}{\alpha_i^\prime(0)}, \quad \forall i\in\{1,
\cdots, I\}\}$   
extends to a  meromorphic map on the whole complex plane with   poles located on a countable
set of affine 
hyperplanes
 $$ z_{\tau(1)}+\cdots+
z_{\tau(i)}\in -\frac{-a_{\tau(1)}-\cdots-a_{\tau(i)}+\lambda_{\tau,i}+\N_0}{q}
,i\in \{1, \cdots, I\}, 
\quad \tau\in \Sigma_I,$$ 
and where $  \lambda_{\tau,i}\in [-n\, i, 0[\cap \Z$  depends on the matrix
$B$. 
\end{thm}
\begin{rk}An immediate but important consequence of this theorem is the fact
  that 
if none of  the partial sums of the orders $a_i$ are  integers then the
hyerplanes of poles of the map $\underline z\mapsto
\int_{\R^{nL}}\widetilde {\cal R}(\tilde \sigma)(\underline z)\circ B$ do not
contain $0$ so that the map is holomorphic in a neighborhood of $0$.
\end{rk}
Before going to the proof, let us illustrate this result by an example.  
\begin{ex}If we choose $I= 3, L=2$, $\sigma_i, i=1,2, 3$ , ${\cal
    R}(\sigma)(z)(\xi)= \sigma(\xi) \, \langle \xi\rangle^{-z}$ (here $q=1$) with $\langle
  \xi\rangle:= \sqrt{1+\vert \xi\vert^2}$ and $B$ as in
  Example \ref{ex:Feynmanmatrix}, this yields back the known fact that the map 
$(z_1, z_2, z_3) \mapsto \int_{\R^{n2}}\frac{1}{(\vert \xi_1\vert^2+ 1)^{a_1-z_1}}
\frac{1}{(\vert\xi_1+ \xi_2\vert^2+ 1)^{a_2-z_2}}  \frac{1}{(\vert
  \xi_2\vert^2+ 1)^{a_3-z_3}}\,d\xi_1\, d\xi_2 $  has a meromorphic extension to the plane with
poles on hyperplanes defined by equations involving partial sums of the
$z_i$'s. Whenever $a_1, a_2, a_3, a_1+a_2, a_2+ a_3, a_1+
a_3, a_1+a_2+a_3$ are not integers, the map is holomorphic in a neighborhood
of $0$.
\end{ex}
Setting $z_i=z$ in the above theorem leads to the following result.
\begin{cor}  \label{cor:meromultintsymb}Let ${\cal R}: \sigma\mapsto \sigma(z)$ be a holomorphic
  regularisation procedure on $CS(\R_+)$ and let 
$\xi\mapsto \sigma_i(\xi):= \tau_i(\vert
\xi\vert)\in CS(\R^n), i=1, \cdots, I$ be  {\rm radial} polyhomogeneous symbols    of order $a_i$ which are
sent via ${\cal R}$ to  $\xi\mapsto \sigma_i(z)(\xi):= {\cal R}(\tau)(z)(\vert \xi\vert)$ of non constant affine order
$\alpha_i(z)=-qz_i+a_i$, for some positive real number $q$. For any  matrix $B$ of siwe $I\times L$ and rank $L$,   the map 
$$z\mapsto
\int_{\R^{nL}} \left({\cal R} (\sigma_1)(z)\otimes \cdots \otimes {\cal R} (\sigma_I)(z)\right)\circ B$$
which is well defined and  holomorphic on the domain  $D=\{ z\in \C,\quad {\rm
   Re}(z) > -\frac{a_i+n}{\alpha_i^\prime(0)}, \quad \forall i\in\{1,
\cdots, I\}\}$   
extends to a  meromorphic map on the whole complex plane with  a countable set
of poles with finite multiplicity
 $$ z\in -\frac{-a_{\tau(1)}-\cdots-a_{\tau(i)}+\lambda_{\tau,i}+\N_0}{q\, i}
,i\in \{1, \cdots, I\}, 
\quad \tau\in \Sigma_I,$$ where as before $\lambda_{\tau,i}\in [-n\, i, 0[$ is an integer
depending on the matrix $B$. 
\end{cor} 

In order to prove Theorem \ref{thm:meromultintsymb}, we proceed in several steps, first reducing
the problem to step matrices $B$, then to symbols of the type  $\sigma_i: \xi \mapsto
  (\xi^2+1)^{a_i}$ and finally proving the meromorphicity for such symbols and matrices. 
\subsection*{Step 1: Reduction to step matrices}
We call an $I\times J$ matrix $B$  with real
 coefficients a step matrix  if it fulfills the following condition
  \begin{equation}\label{eq:matrix} \exists i_1< \cdots< i_L \quad {\rm in}
    \quad  \{1,
    \cdots,  I\}\quad
    {\rm s.t}\quad  b_{i l}=0 \quad {\rm if} \quad 
i>i_l\quad {\rm and} \quad  b_{i_l, l}\neq 0.
\end{equation}
\begin{rk}This condition actually says  that the matrix has rank $ \geq L$. It
  $J=L$ then it has rank $L$.
\end{rk}
\begin{prop}If Theorem \ref{thm:meromultintsymb} holds for  step
  matrices  then it holds for any matrix $B$.
\end{prop}
{\bf Proof:} \begin{itemize}
\item Let us first observe that if the result holds for a matrix $B$
then it holds for any matrix $P\, B\, Q$ where $P $ and $Q$ are permutation
matrices i.e. up to a relabelling of the symbols and the variables. Indeed,  a  permutation $\tau\in \Sigma_I$  on the lines induced by the matrix $P$
amounts to a relabelling of the symbols; since the statement should hold for
all radial symbols, if it holds for $\tilde \sigma= \sigma_1\otimes \cdots
\otimes \sigma_I$ then it also holds for $\sigma_{\tau(1)}\otimes \cdots \otimes
\sigma_{\tau(I) }$. Hence, if the statement of the theorem holds for a matrix $B$ it also
holds for the matrix $P\,B$.\\
 Assuming the statement of the theorem holds for a matrix $B$, then it  also
 holds for the matrix $B\, Q$. Indeed,    a permutation $\tau\in \Sigma_L$ on the
columns induced by the matrix $Q$ amounts to a relabelling of the variables
$\xi_l$. By Proposition \ref{prop:holmultintsymb} we know that if $B$ has rank
$L$ then both 
the maps $\underline z \mapsto
\int_{\R^{nL}}\widetilde {\cal R}(\tilde \sigma)(\underline z)\circ B$ and
   $\underline z \mapsto
\int_{\R^{nL}}\widetilde {\cal R}(\tilde \sigma)(\underline z)\circ B\, Q$ are
well defined and holomorphic on the domain
  $D=\{\underline z\in \C^I,\quad {\rm
   Re}(z_i) > -\frac{a_i+n}{\alpha_i^\prime(0)}, \quad\forall i\in\{1, \cdots,
 I\}\}$. By the Fubini property
 we further have that 
$$\int_{\R^{nL}}\widetilde {\cal R}(\tilde \sigma)(\underline z)\circ
B=\vert {\rm det} Q\vert\, \int_{\R^{nL}}\widetilde {\cal R}(\tilde \sigma)(\underline z)\circ B\, Q\quad
\forall \underline z\in D.$$
If by assumption, the r.h.s has a meromorphic extension $\underline z\mapsto
\cutoffint_{\R^{nL}}\widetilde {\cal R}(\tilde \sigma)(\underline z)\circ B\, Q$  then so does the
l.h.s. have a meromorphic extension 
$$\cutoffint_{\R^{nL}}\widetilde {\cal R}(\tilde \sigma)(\underline z)\circ
B:=\vert {\rm det} Q\vert \, \cutoffint_{\R^{nL}}\widetilde {\cal R}(\tilde
\sigma)(\underline z)\circ B\, Q$$
which moreover has  the same pole structure.
\item 
Let $B$ be a  non zero matrix. Then  there is an invertible matrix
$P$ and step matrix $T$
such that $P\, B^t= T$ where $B^t$ stands for the transpose of $B$. Hence the
existence of an invertible  matrix $Q=\left( P^t\right)^{-1}$ such that $B=
T^t \, Q$. If $B$ has rank $L$ then so does the matrix $T^t$; along  the same
lines as above, one shows that if the statement of the theorem holds for $T^t$
then it holds for $B$. On the other hand, there are permutation matrices $P$
and $Q$ such that $S:=P \, T^t\, Q$ is a step matrix  for the transpose of a step matrix  can
be turned into a step matrix by iterated permutations on its lines and columns. If
the theorem holds for step matrices then by the first part of the proof, it
also holds for $T^t$ and hence for $B$.\end{itemize} \endsquare

\subsection*{Step 2: Reduction to symbols $\sigma_i: \xi \mapsto
  (\xi^2+1)^{a_i}$}
Let us first describe the asymptotic behaviour of  classical  radial symbols. 
\begin{lem} \label{lem:radial}Given a {\rm radial}  polyhomogeneous  symbol $ 
\sigma: \xi \mapsto
  \tau(\vert \xi\vert)$ on $\R^n$, $\tau\in CS(\R_+)$ of  order
  $a$ there are real numbers $c_j, j\in \N_0$ such that
 $$\sigma( \xi)\sim \sum_{j=0}^\infty c_j\, \langle
  \xi\rangle^{a-j}$$ 
where $\sim$  stands for the equivalence of symbols modulo smoothing symbols.  Here, as before we have set
$\langle \xi\rangle= \sqrt{1+\vert \xi\vert^2}$. 
\end{lem}
{\bf Proof:} A radial  polyhomogeneous  symbol $\sigma$ on $\R^n$ of  order
  $a$ can be written $$\sigma(\ \xi)= \sum_{j=0}^{N-1}  \tau_{
  a-j}\left(\vert \xi \vert\right)\, \chi(\vert \xi\vert)+
\tau^{ (N)}(\vert \xi\vert)$$ where $N$ is a positive integer,
$\tau^{ (N)}$ is a polyhomogeneous symbol the  
order of which has real part  no larger than ${\rm Re}( a)-N$ and where  $\tau_{a-j}$ are positively
homogeneous functions of degree $a-j$. $\chi$ is a smooth cut-off
function on $\R_0^+$ which  vanishes in a small neighborhood of $1$ and is
identically  $1$ outside the unit interval. Setting
$\gamma_{a-j}:=\tau_{a-j}(1)$  we write
\begin{eqnarray*}
 \tau_{a-j}\left(\vert \xi\vert \right)\, \chi(\vert \xi\vert)&=& \gamma_{a-j}\,
\vert \xi\vert^{a-j} \, \chi(\vert \xi\vert) \\
&=& \gamma_{a-j}\,
(\langle \xi\rangle^2-1)^{\frac{a-j}{2}} \, \chi(\vert \xi\vert) \\
&=& \gamma_{a-j}\,\langle \xi\rangle^{a-j}\,
(1-\langle\xi\rangle^{-2})^{\frac{a-j}{2}} \, \chi(\vert \xi\vert) \\
&\sim&  \gamma_{a-j}\,\langle \xi\rangle^{a-j}\, \chi(\vert \xi\vert)\,
\sum_{k_j=0}^\infty b_{k_j}\langle \xi\rangle^{-2\,k_j}   \\
&\sim& 
\sum_{k_j=0}^\infty c_{k_j}\langle \xi\rangle^{a-j-2k_j}   
\end{eqnarray*}where we have set $c_{k_j}:= \gamma_{a-j}\,b_{k_j}$ for some sequence $b_{j_k}, k\in \N_0$ of real numbers
depending on $a$ and $j$ 
 and used the fact that
$\chi\sim 1$.  Applying this to each $\tau_{a-j}$ yields for any
positive integer $N$, the existence
of a symbol $\tilde \tau^{(N)}(\vert \xi\vert)$  the  order of which has real part no
larger than ${\rm Re}(a)-N$ and constants $\tilde c_j$  such that
 $$\sigma( \xi)= \sum_{j=0}^{N-1}  \tilde c_{j}\langle \xi\rangle^{a-j}+
\tilde \tau^{ (N)}(\vert \xi\vert)$$
which ends the proof of the lemma.
\endsquare\\ \\
 Let $\xi\mapsto \sigma_1(\xi):=  \tau_1(\vert \xi\vert), \cdots,\xi\mapsto 
\sigma_I(\xi):=  \tau_I(\vert \xi\vert)$
  be  radial  polyhomogeneous  symbol on $\R^n$ of  order $a_1, \cdots,
  a_I$ respectively which  we write
  \begin{eqnarray}\label{eq:sigmairadial}\sigma_i(  x_i)&=& \sum_{j_i=0}^{N_i-1} 
    \tau_{i, a_i-j_i}(\vert  x_i\vert)+
\tau_i^{ (N_i)}( \vert x_i\vert)\, \chi(\vert x_i\vert)\nonumber \\
&=& \sum_{j_i=0}^{N_i-1} c^i_{j_i} \,\langle x_i\rangle^{a_i-j_i}+
\widetilde \tau_i^{ (N_i)}(\vert x_i\vert)
\end{eqnarray} where $N_i, i=1, \cdots, I$ are positive integers, $\tau_{i,
  a_i-j_i}, i=1, \cdots, I$ are homogeneous functions of degree
$a_i-j_i$, 
$\tau_{i}^{ (N_i)},\widetilde \tau_{i}^{ (N_i)},  i=1, \cdots, I$   polyhomogeneous symbols of 
order with real part no larger than $ a_i-N_i$ and where we have set  $c^i_{j_i}:=\tau_{i, a_i-j_i}(1), i=1, \cdots, I$.\\
It follows
that   
\begin{equation}\label{eq:asympttensorradial}\prod_{i=1}^I \sigma_i(\xi_i)=  \lim_{N\to \infty}
\sum_{j_1=0}^{N-1}\cdots \sum_{j_I=0}^{N-1} c^1_{j_1}\cdots  c^I_{j_I} \,\langle
\xi_1\rangle^{a_1-j_1}\cdots \langle
\xi_I\rangle^{a_I-j_I}
\end{equation}
in the Fr\'echet  topology on symbols of constant order \footnote{This
  Fr\'echet topology was described in a footnote in Section 1.}.\\
\begin{prop} If Theorem \ref{thm:meromultintsymb} holds for  symbols $\sigma_i:
  \xi\mapsto \langle \xi\rangle^{a_i}$ then
  it holds for all classical radial symbols.
\end{prop}
{\bf Proof:}
 Let  $B$ be an $L\times I$  matrix of rank $L$  and let $
\sigma_1, \cdots,
  \sigma_I$ be {\rm radial} polyhomogeneous symbols in $CS(\R^n)$ with orders
  $a_1,\cdots , a_I$ respectively. For each $j_i\in \N, i\in \{1, \cdots, I\}$
  we set $ \rho^{j_i}_i(\xi):= \langle \xi\rangle^{a_i-j_i}
$ and for all multiindices $(j_1, \cdots, j_I)$ we set $\tilde \rho^{j_1\cdots
  j_I}:= \otimes_{i=1}^I \rho^{j_i}_i$.  \\  Let us first observe that since
${\rm Re}(a_i)-j_i\leq {\rm Re}(a_i)$, the maps
$$\underline z\mapsto
\int_{\R^{nL}}\widetilde {\cal R}(\tilde \rho^{j_1\cdots j_I})(\underline z)\circ B$$
 are all well defined and  holomorphic on  the domain  $D=\{\underline z\in \C^I,\quad {\rm
   Re}(z_i) > -\frac{a_i+n}{\alpha_i^\prime(0)}, \quad \forall i\in \{1,
\cdots, I\}$. \\ Let us assume that the theorem holds for this specific class of
symbols. Then using again the fact that  $\rho_i^{j_i}$ has order  $a_i-j_i$ which
differs from $a_i$ by a non negative  integer, and replacing $a_i$ by $\alpha_i(z_i)$, it follows that these maps 
extend to   meromorphic maps $$\underline z\mapsto
\cutoffint_{\R^{nL}}\widetilde {\cal R}(\tilde \rho^{j_1\cdots
  j_I})(\underline z)\circ B$$ on the whole complex plane with   poles
$\underline z=(z_1, \cdots, z_I)$ such that 
 $$\alpha_{\tau(1)}( z_{\tau(1)})+\cdots+
\alpha_{\tau(i)}(z_{\tau(i)})\in \lambda_{\tau,i}+\N_0,  \quad \tau\in \Sigma_I,$$ or equivalently with poles
located on a  countable
set of affine 
hyperplanes$$ z_{\tau(1)}+\cdots+
z_{\tau(i)}\in -\frac{-a_{\tau(1)}-\cdots -a_{\tau(i)}+
  \lambda_{\tau,i}+\N_0}{q}, \quad \tau\in \Sigma_I,$$
with $\lambda_{\tau, i}\in [-n\, i, 0[\cap \Z$  depending on the matrix $B$.\\

Then by (\ref{eq:asympttensorradial}) so does the map
$$\underline z\mapsto
\int_{\R^{nL}}\widetilde {\cal R}(\tilde \sigma)(\underline z)\circ B$$ extend
to a meromorphic map on the complex plane: 
 \begin{eqnarray*}
\underline z&\mapsto&
\cutoffint_{\R^{nL}}\widetilde {\cal R}(\tilde \rho^{j_1\cdots
  j_I})(\underline z)\circ B\\
&:= &\lim_{N\to \infty} \sum_{j_1=0}^{N-1}\cdots \sum_{j_I=0}^{N-1}c^1_{j_1}\cdots
c^I_{j_I} \cutoffint_{\R^{nL}} \prod_{i=1}^I c_{j_1}^1\cdots c_{j_I}^I
\left({\cal R}(\rho^{j_1}_1)(z_1) \cdots{\cal R}(\rho^{j_I}_I)(z_I)
\right)\circ B
\end{eqnarray*} with the same pole structure. Note that for large enough
$j_i$'s, the hyperplanes of
poles do not contain  the origin and that their distance to the origin then
increases as the $j_i$'s further increase.
\endsquare
\subsection*{Step 3: The case of symbols $\sigma_i:\xi\mapsto (\vert \xi\vert^2+1)^{a_i}$  and step matrices}

We are therefore left to prove the statement of the theorem for an  $I\times L$ matrix $B$  with real
 coefficients which  fulfills  condition (\ref{eq:matrix})
and symbols $\sigma_i: \xi\mapsto (\vert \xi\vert^2+1)^{a_i}$. As previously
observed, such a matrix has rank $L$.
\begin{lem} Under  assumption (\ref{eq:matrix})  on
$B=(b_{il})$ the matrix $B^*B$ is positive definite. Note that with the
notations of (\ref{eq:matrix}), we have   $i_l\geq l$. 
\end{lem}{\bf Proof:} For $k\in \R^L$ in the kernel of $ B$, we have 
 $
\sum_{l=1}^L b_{il} \xi_l=0$ for any $i=1, \cdots, I$, which applied to  $i=i_L$  yields
$\sum_{l=1}^L b_{i_Ll} \xi_l=0$. But since by assumption $b_{i_Ll}=0$ for $l<L$
only the term $b_{I_LL} \xi_l$ remains which shows that $\xi_l=0$. Proceeding
inductively yields the positivity of $B^*B$. \endsquare
\begin{prop}\label{prop:meroextmultint} Let  $B:=(b_{il})_{i=1, \cdots, I; l=1, \cdots, L}$ be a matrix  with property (\ref{eq:matrix}).
The map  
$$
(a_1, \cdots, a_I)\mapsto \int_{\left(\R^n\right)^L}
 \prod_{i=1}^I
\langle \sum_{l=1}^L b_{il} \, \xi_l \rangle^{a_i}\, d\xi_1\cdots d\xi_L ,$$ 
which is   holomorphic
 on the domain $D:=\{\underline
 a=(a_1, \cdots, a_I)\in \C^I, {\rm Re}(a_i)<-n, \forall i\in
 \{1, \cdots, I\}\}$, 
has a meromorphic extension  to the complex plane
\begin{eqnarray}\label{eq:meroextmultint}
&&(a_1, \cdots, a_I)\mapsto \cutoffint_{\left(\R^n\right)^L}
 \prod_{i=1}^I
\langle \sum_{l=1}^L b_{il} \, \xi_L \rangle^{a_i}\, d\xi_1\cdots d\xi_L  \\ 
&:=&
\frac{1}{\prod_{i=1}^I \Gamma(-a_i/2)}\,
\sum_{\tau\in \Sigma_I} \frac{ H_{\tau, \underline m}(a_1, \cdots,a_I)}{ \prod_{i=1}^I\left[(a_{\tau(1)}+\cdots+
  a_{\tau(i)}+n\,s_{\tau, i})\cdots (a_{\tau(1)}+\cdots+
  a_{\tau(i)}+n\,s_{ \tau, i} -2m_i )\right]}\nonumber
\end{eqnarray} for some holomorphic map  $H_{\tau, \underline m}$ on the
domain  $\cap_{i=1}^I\{{\rm Re}(a_{\tau(1)}+\cdots+
  a_{\tau(i)})+2m_i<-n\,s_{\tau, i}\}$,  with $\tau\in \Sigma_I$ and  
$\underline m:= (m_1, \cdots, m_{I})$ a multiindex of  non negative
integers. The 
 $s_{\tau,i}\leq i$'s are positive integers which depend 
on the permutation $\tau$, on the size  $L\times I$ and  shape (i.e. on the $l_i$'s)  of the
matrix but not on the actual
coefficients of the matrix.\\  The poles of this meromorphic extension 
lie  on a  countable set of affine
hyperplanes $a_{\tau(1)}+\cdots
+a_{\tau(i)}\in \lambda_{\tau,i}+\N_0$ with
$\tau\in\Sigma_I, \quad i\in \{1, \cdots, I\}$, $\lambda_{\tau,i}:= -n
\,s_{\tau, i}\in [-n\, i,  0[\cap \Z$.
\end{prop}
The proof, which is rather technical and lengthy  is postponed to
 the Appendix.  It closely follows Speer's proof \cite{Sp} which  uses  iterated Mellin transforms
and integrations by parts.

 \section{Renormalised multiple   integrals  with constraints}
Let us consider the set
$${\cal A}_I:=\{(\sigma_1\otimes \cdots\otimes \sigma_I)\circ B, \quad
\sigma_i\in CS_{\rm rad}(\R^n),\quad  B\in {\cal
M}_{I,L}(\R),\quad {\rm rk}B=L, \quad L\in \N\},$$  where  $CS_{\rm rad}(\R^n)$ stands for
the algebra of classical radial symbols $\xi\mapsto \tau(\vert \xi\vert)$ with
$\tau\in CS(\R_+)$, ${\cal M}_{IL}(\R)$ for the set of matrices of size
$I\times L$ with coefficients in $\R$. 
The map \begin{eqnarray*}{\cal A}_I\times {\cal A}_{I^\prime}&\to&{\cal A}_{I+I^\prime}\\
\left(\tilde\sigma \circ B\right)\times\left(\tilde \sigma^\prime\circ
  B^\prime\right)&\mapsto &\left(\tilde\sigma \circ B\right)\bullet \left(\tilde \sigma^\prime\circ
  B^\prime\right):= \left(\sigma\otimes \sigma^\prime\right)\circ
(B\oplus B^\prime),
\end{eqnarray*}
where $\oplus$ stands for the  Whitney sum: 
$$B\oplus B^\prime:=  \left( \begin{array}{llc} &B &0 \\
                                  & 0 & B^\prime\\ \end{array}\right) $$
induces  a morphism of filtered algebras on ${\cal A}:=\cup_{I=1 }^\infty
{\cal A }_I$.  \\ Let us also  introduce the set \begin{eqnarray}\label{eq:BI}
{\cal B}_I&:= &\{f: \C^I\mapsto \C, \quad {\rm s.t}\quad \exists (m_1, \cdots,
m_I)\in \N_0^I, \quad  {\rm the}\quad {\rm
  map}\nonumber\\
&&\quad (z_1,\cdots,
z_I)\mapsto \quad f(z_1, \cdots, z_I)\,\prod_{\tau\in \Sigma_I}\,\left(\prod_{i=1}^I (z_{\tau(1)}+\cdots+z_{\tau(i)})^{m_i}\right)\nonumber\\
&\quad& {\rm is}\quad {\rm holomorphic}\quad {\rm around }\quad
\underline z=0\},
\end{eqnarray}
then ${\cal B}:= \bigcup_{I=1}^\infty{\cal B}_I$  is a filtered algebra for
the ordinary product of functions. 
\\ \\
 The following proposition is an easy consequence of
Theorem
 \ref{thm:meromultintsymb}.
\begin{prop}Let ${\cal R}: \sigma\mapsto \sigma(z)$ be a holomorphic
  regularisation procedure on $CS(\R_+)$ which sends a symbol $\tau$ of order $a$to   ${\cal R}(\tau)(z)$ of non constant affine order
$-qz_i+a$, for some positive real number $q$. 
The map: 
\begin{eqnarray*}\Phi^{{\cal R}}: {\cal A}&\to& {\cal B}\\
\left(\tilde\sigma \circ B\right)&\mapsto
&\left(\underline z\mapsto \cutoffint_{\R^{nL}}\widetilde {\cal R}\left(\tilde\sigma \right)(\underline z)\circ B\right),
\end{eqnarray*}is a morphism of algebras.
\end{prop} 
{\bf Proof:} It follows from Theorem
 \ref{thm:meromultintsymb} that if $\tilde\sigma\circ B$ lies in ${\cal A}_I$
 then  the map $\underline z\mapsto
 \cutoffint_{\R^{nL}}\widetilde {\cal R}\left(\tilde\sigma \right)(\underline
 z)\circ B$ lies in  ${\cal B}_I$. The factorisation property w.r.to the
 product $\bullet$:$$\Phi^{\cal R}\left[\left(\tilde\sigma \circ B\right)\bullet\left(\tilde \sigma^\prime\circ
  B^\prime\right)\right]= \Phi^{\cal R}\left(\tilde\sigma \circ
  B\right)\, \cdot \, \Phi^{\cal R} \left(\tilde \sigma^\prime\circ
  B^\prime\right)$$
 then follows by analytic continuation  from the corresponding factorisation property on the domain of
 holomorphicity. \endsquare
\subsection{Renormalisation via generalised evaluators}
With the help of the  morphism $\Phi^{\cal R}$, we now build a character
$\phi^{\cal R}: {\cal A}\to \C$ 
which boils down to building renormalised integrals which  factorise on
disjoint sets of constraints.
Generalised evaluators (see e.g.\cite{Sp}) at $0$  provide an adequate procedure to
extract "multiplicative" finite parts at $0$ of meromorphic functions in a filtered
algebra of the  type ${\cal F}=\cup_{k=1}^\infty{\cal F}_I$ with:
\begin{eqnarray}\label{eq:FI}
{\cal F}_I&:= &\{f: \C^k\mapsto \C, \quad \exists \, (m_1, \cdots,
m_I)\in \N_0^I,
 \nonumber\\ && {\rm s.t} \quad{\rm the}\quad {\rm map}\quad (z_1,\cdots,
z_I)\mapsto \quad f(z_1, \cdots, z_I)\,\prod_{\tau\in \Sigma_I}\,
\prod_{i=1}^I 
(L_i^k(z_{\tau(1)}, \cdots,
z_{\tau(I)}))^{m_i}\nonumber\\
&& \quad  {\rm is}\quad {\rm holomorphic}\quad {\rm around }\quad
\underline z=0\}
\end{eqnarray}
where  $ L_1^I, \cdots, L_I^I$ are  linear forms 
$L_i^I(\underline z)=\sum_{j=1}^I a_{ij}^I z_j, i\in \{1, \cdots, I\}$ such that the matrix
$(a_{ij}^I)$ can be embedded in the upper left corner of the matrix
$(a_{ij}^{I+1})$.
\begin{ex} ${\cal B}=\cup_{I\in \N}{\cal B}_I $ with ${\cal B}_I $ defined in
  (\ref{eq:BI})
 is such a filtered algebra with $ L_i^I(\underline z)= z_1+\cdots+z_i$. 
\end{ex}
\begin{rk}Such filtered algebras are stable under  holomorphic  reparametrisations $z\mapsto
\kappa(z)$ such that $\kappa(0)=0$ and $\kappa^\prime(0)\neq 0$, i.e.
$$f_I \in {\cal F}_I\Rightarrow f_I\circ \kappa^{\otimes I}\in {\cal
  F}_I.$$ Indeed,   the pole  part  $\frac{1}{ L(z_1, \cdots, z_I)}$
corresponding to a
linear form 
$L(z_1, \cdots, z_I)=\sum_{i=1}^I a_i\, z_i$ transforms to 
\begin{eqnarray*}
\frac{1}{a_1\kappa(z_1)+\cdots
  +a_I\kappa(z_I)}&&=\frac{1}{\kappa^\prime(0)\left[a_1\,(z_1+\frac{\kappa^{\prime\prime}(0)}{2 \, \kappa^\prime(0)} z_1^2+o(z_1^2))+\cdots +a_I\,(z_I+\frac{\kappa^{\prime\prime}(0)}{2\,\kappa^\prime(0) } z_I^2+o(z_I^2))\right]}\\
&&=\frac{1}{\kappa^\prime(0)\,L(z_1, \cdots, z_I)}\frac{1}{1+
  \frac{\kappa^{\prime \prime}(0)}{2\, \kappa^\prime(0)}\,\frac{ \sum_{i=1}^I
    (z_i^2+ o(z_i^2))}{L(z_1, \cdots, z_I)}}\\
&=& \frac{1}{\kappa^\prime(0)\,L(z_1, \cdots, z_I)}\left(
  1-\frac{\kappa^{\prime \prime}(0) \sum_{i=1}^I
    (z_i^2+ o(z_i^2))}{2\,\kappa^\prime(0)\, L(z_1, \cdots, z_I)}\right)
\end{eqnarray*}
which is a meromorphic map with poles of the same type. 
\end{rk}
\begin{defn} (see e.g. \cite{Sp}) A generalised  evaluator at $0$  on the filtered algebra ${\cal
    F}=\bigcup_{i=1}^\infty {\cal F}_I$ 
 is a family of maps ${\cal E}=\{{\cal E}_I,
  I\in \N\}$, ${\cal E}_I: {\cal F}_I\to \C$ such that 
\begin{enumerate}
\item ${\cal E}$ is linear,
\item ${\cal E}$ coincides with the evaluation at $0$ on analytic functions
  around $0$, 
\item ${\cal E}$ is continuous for the uniform convergence of analytic
  functions,
\item ${\cal E}$ is symmetric in the variables $z_i$'s, 
\item ${\cal E}$ is compatible with the filtration on ${\cal F}$,
\item ${\cal E}$ is multiplicative on tensor products:
\begin{equation}\label{eq:multev}{\cal E}_{I+I^\prime}(f\otimes f^\prime)= {\cal
    E}_I(f)\, {\cal E}_{I^\prime}( f^\prime)
\end{equation}
for any $f\in {\cal F}_I$ depending only on the   first $I$ variables $z_1, \cdots, z_I$,
$f^\prime\in {\cal F}_{I^\prime}$ on the remaining $I^\prime$ variables  $z_{I+1}, \cdots, z_{I+I^\prime}$.
\end{enumerate}
\end{defn}
The map ${\cal E}$ on ${\cal F}$ defined on ${\cal F}_I$ by:  \label{ex:evaluator} \begin{equation}
{\cal E}^0_I(f):=\frac{1}{I!}  \sum_{\tau\in \Sigma_I}{\rm
  fp}_{z_{\tau(1)}=0}
\left(\cdots
    \left({\rm fp}_{z_{\tau(I)}=0} \, f(\underline z)  \right)\cdots\right) 
\end{equation}
yields a generalized evaluator   ${\cal E}^0$ at $0$ on ${\cal F}$.\\
A holomorphic reparametrization $z\mapsto \kappa(z)$ such that $\kappa(0)=0$
and $\kappa^\prime(0)\neq 0$
 induces another evaluator ${\cal E}^\kappa$ defined on ${\cal F}_k$ by
$$ {\cal
  E}_I^{\kappa}(f):= {\cal E}_I^0(f_I\circ \kappa^{\otimes I})$$
since $f_k \in {\cal F}_I\Rightarrow f_I\circ \kappa^{\otimes I}\in {\cal
  F}_k$ by the above remark.\\ In general, ${\cal E}^\kappa\neq
{\cal E}^0$ as the following example shows. 
 \begin{ex}Note that $${\cal F}_1=\{f: \C\to \C,\quad \exists m\in \N_0,\quad {\rm
     s.t}\quad z\mapsto f(z) \, z^m\quad {\rm is}\quad {\rm holomorphic}\quad
 {\rm around }\quad 0\}$$ corresponds to  functions in one
   variable meromorphic in a
 neighborhood of $0$   with poles at $z=0$.  The evaluator ${\cal E}^0$ on ${\cal
    F}_1$ applied to $f(z)=\sum_{i=1}^m \frac{a_i}{z^i}+o(z)$ singles out the
  finite part  ${\cal
    E}_1^0(f)={\rm fp}_{z=0} f(z):=\lim_{z\to 0}\left(f(z)-\sum_{i=1}^m
    \frac{a_i}{z^i}\right) $. When applied to $f\circ \kappa(z)$ it picks up extra contributions since
\begin{eqnarray*}\frac{1}{\left(\kappa(z)\right)^i}&= &\frac{1}{\left(\kappa^\prime(0) z+ \frac{\kappa^{\prime
      \prime}(0)}{2}z^2+ o(z^2)\right)^i}\\
&=&\frac{1}{\left(\kappa^\prime(0)\right)^i z^i \left(1 +  \frac{\kappa^{\prime
      \prime}(0)}{2\kappa^\prime(0)}z+ o(z)\right)^i}\\
&=&\frac{ 1 +\sum_{j=1}^J \alpha^i_j \,z^j+
  o(z^J)}{\left(\kappa^\prime(0)\right)^i z^i}\quad {\rm for}\quad{\rm
  some}\quad (\alpha^i_1, \cdots, \alpha^i_J)\in \C^J\\
&\Longrightarrow & {\cal E}_1^0 \left(\frac{1}{\left(\kappa(z)\right)^i}\right)=
\delta_{ij}\, \frac{\alpha^i_j }{\left(\kappa^\prime(0)\right)^i}
\end{eqnarray*}
which in turn implies that ${\cal E}_1^0(f\circ \kappa)= {\cal
  E}_1^0(f)+\delta_{ij} \frac{a_i\, \alpha^i_j
}{\left(\kappa^\prime(0)\right)^i}  $.
\end{ex}
A change of variable $T_k: (z_1, \cdots, z_k)\mapsto T_k ( z_1, \cdots, 
z_k)$ with ${\cal T} =\{T_k\in {\rm GL}_k(\C), k\in \N\}$ a family of matrices nested in one
another i.e. such that the matrix $T_k$ can be
embedded in the upper left  corner of the matrix $T_{k+1}$,  gives rise to
another evaluator ${\cal E}^{\cal T}$ defined on ${\cal F}_k$ by  $$ {\cal
  E}_k^{\cal T}(f):= {\cal E}_k^0(f_k\circ T_k).$$
\begin{rk} Clearly, $ {\cal
  E}_1^{\cal T}(f):= {\cal E}_1^0(f_1\circ T_1)$ for any $ f\in {\cal F}$ since the finite part at
$z=0$ of a
meromorphic function  $z\mapsto f(z)$ around zero is insensitive to a linear
transformation $z\mapsto a\, z$ with $a\neq 0$. 
 \end{rk}
However, in general, ${\cal E}^{\cal T}\neq
{\cal E}^0$ as the following example shows. 
\begin{ex} The evaluator
  ${\cal E}^0$ at $0$ on ${\cal B}_2$ applied to the  map  $f:
  (z_1,z_2)\mapsto \frac{z_1+z_2}{z_1}$  yields:
$${\cal E}_2^0(f)= \frac{
{\rm fp}_{z_2=0}\left(  {\rm fp}_{z_1=0}\left(1+ \frac{z_2}{z_1}\right) \right)+ {\rm
  fp}_{z_1=0} \left({\rm fp}_{z_2=0}\left(1+ \frac{z_2}{z_1}\right)\right)}{2}= 1.$$  In contrast, the evaluator
  ${\cal E}^{\cal T}$ corresponding  to maps $T_k(z_1,z_2, \cdots, z_k)=
  (z_1,z_2-z_1,\cdots
  ,z_k-z_{k-1})$ yields $${\cal E}_2^{\cal T}(f)= {\cal E}_2^0(f\circ T_2(f))=\frac{
{\rm fp}_{u_2=0}\left(  {\rm fp}_{u_1=0}\frac{u_2}{u_1} \right)+ {\rm
  fp}_{u_1=0} \left({\rm fp}_{u_2=0}\frac{u_2}{u_1}\right)}{2}=0.$$
\end{ex}
Combining the two types of transformations on evaluators, yields a family 
${\cal E}^{\kappa, {\cal T}}$ of evaluators defined on ${\cal F}_k$ by: 
 $$ {\cal
  E}_k^{\kappa, {\cal T}}(f):= {\cal E}_k^0(f_k\circ T_k\circ \kappa).$$
\begin{rk} This raises the question whether  such evaluators linearly span all
  evaluators. 
\end{rk}
\begin{thm}\label{thm:evaluatorsrenormalised}Let ${\cal R}: \sigma\mapsto \sigma(z)$ be a holomorphic
  regularisation procedure on $CS(\R_+)$ which sends a symbol $\tau$ of order
  $a$ to   ${\cal R}(\tau)(z)$ of non constant affine order
$-qz_i+a$, for some positive real number $q$ and let ${\cal E}$ be a
generalised 
evaluator at $0$  on  the algebra $ {\cal B}$ of meromorphic maps then
the map $\phi^{{\cal R}, {\cal E}}:= {\cal E}\circ \Phi^{{\cal R}}$: 
\begin{eqnarray*}\phi^{{\cal R}, {\cal E}}: {\cal A}&\to& \C\\
\tilde\sigma \circ B&\mapsto
&{\cal E}\circ\cutoffint^{{\cal R}, {\cal E}}_{\R^{nL}}\tilde\sigma
\circ B:= {\cal E}\circ\cutoffint_{\R^{nL}}\widetilde {\cal R}\left(\tilde\sigma \right)\circ B,
\end{eqnarray*}is a character. Whenever $\tilde \sigma=\sigma_1\otimes \cdots
\otimes \sigma_I$ with $\sigma_i$ of order $a_i$ with real part $<-n$ then
${\cal E}\circ\cutoffint_{\R^{nL}}\widetilde {\cal R}\left(\tilde\sigma
\right)\circ B$  coincides with the ordinary integral 
$\int_{\R^{nL}}\tilde\sigma \circ B$.
\end{thm} 
{\bf Proof:} The multiplicativity easily  follows from combining the multiplicative properties
of the morphism 
$\Phi^{\cal R}$ and the evaluator ${\cal E}$. The fact that it  coincides with the ordinary integral 
$\int_{\R^{nL}}\tilde\sigma\circ B$ when $\sigma_i$ has order $a_i$
with real part $<-n$, follows from the fact that the map $\Phi^{\cal R}$ is
then holomorphic around $0$ combined with the fact that  evaluators at
$\underline z_0$  on holomorphic
functions around a point $\underline z_0$ indeed boil down to  evaluating the
function at the point $\underline z_0$. \endsquare
\subsection{Renormalisation via Birkhoff factorisation} We now  give an
alternative renormalisation procedure for multiple  integrals
of symbols with linear constraints in the case of  equal symbols
$\sigma_i=\sigma$ with $\sigma$ some 
fixed classical radial symbol. The only freedom left is the choice of the matrix $B$
corresponding to the linear constraints.  Following Connes and Kreimer
\cite{CM}, we carry out this
renormalisation via a  Birkhoff factorisation on a Hopf algebra (here a
Hopf algebra of
matrices plays the role of their Hopf algebra of Feynman diagrams) with the help of   a morphism on this algebra  with values in
meromorphic maps.  \\
We first introduce a Hopf algebra of matrices. 
Let ${\cal H}_L:=\{B\in {\cal M}_{I, L}(\R),\quad {\rm rk}B=L, \quad I\in \N \}$ then ${\cal  H}=\bigcup_{L\in \N} {\cal H}_L$ is
filtered by the rank of the matrix; if $B$ has rank $L$ and $B^\prime$ has
rank $L^\prime$ then $B\oplus B^\prime $ has rank $L+L^\prime$. \\
Writing a matrix $B=(b_{il})_{i=1, \cdots, I; l=1, \cdots, L}$   in terms of its column vectors $B=
[C_1, \cdots, C_L],$ where $C_l= (b_{il})_{i=1, \cdots, I}$, we can
 equip ${\cal H}$ with the following coproduct which boils down to a
 deconcatenation coproduct on column vectors: 
\begin{eqnarray}\label{eq:Delta}\Delta: {\cal H}&\to& {\cal H}\otimes{\cal H}\\
\left[C_1, \cdots, C_l\right] &\mapsto & \sum_{\{l_1, \cdots,l_{p}\}\subset \{1,
  \cdots, L\}}
\left[C_{l_{1}}, \cdots , C_{l_{p}}\right]\bigotimes \left[
  C_{l_{p+1}}, \cdots, C_{l_{p+q}}\right]\nonumber
\end{eqnarray}
   where we have set $L=p+q$ so that  $\{1, \cdots, L\}$ is the disjoint union
of $ \{l_1, \cdots,l_{p}\}$ and $ \{l_{p+1},
\cdots,l_{p+q}\}$. 
\begin{prop}$\left({\cal H}, \oplus, \Delta\right)$ is a graded
  cocommutative Hopf algebra.
\end{prop}
{\bf Proof:} 
We use Sweedler's notations and write in a compact form
$$ \Delta B= \sum_{( B)}B_{(1)} \otimes B_{(2)}.$$
\begin{itemize}
\item The coproduct $\Delta$ is compatible with the filtration
 since it sends ${\cal H}_
L$ to $\bigoplus_{p+q=L} {\cal H}_p\otimes {\cal H}_q$.   
\item The product given by the Whitney sum  $\oplus$  is not  commutative
  since one does not expecto  $B\oplus B^\prime$ to coincide with $ B^\prime
  \oplus B$ for any two  matrices $B$ and $B^\prime$.
\item The product is clearly
associative $ (B\oplus B^\prime)\oplus B^{\prime \prime}= B\oplus
(B^\prime\oplus B^{\prime \prime})$ for any three matrices $B, B^\prime,
B^{\prime \prime}$. 
\item The coproduct  $\Delta$ is clearly  cocommutative since
$\tau_{12} \circ \Delta (B)= \Delta (B)$. 
\item The coproduct  $\Delta$ is coassociative since
\begin{eqnarray*}
(\Delta\otimes 1)\circ \Delta (B)&=&\sum (B_{(1:1)} \otimes B_{(1:2)} )
\otimes B_{(2)}\\
&=& \sum (B_{(1)} \otimes (B_{(2:1)} 
\otimes B_{(2:2)}\\
&=& (1\otimes \Delta) \circ \Delta(B).
\end{eqnarray*}
\item The coproduct  $\Delta$ is compatible with  the  Whitney sum. 
\begin{eqnarray*}
\Delta  \circ m \left(B\otimes B^\prime \right)
&=&  \Delta\left(B\oplus B^\prime \right)\\
&=& \sum_{(B\oplus B^\prime)}(B\oplus B^\prime)_{(1)}  \otimes (B\oplus B^\prime)_{(2)}\\
&=& ( m\otimes m )\circ  \tau_{23} \left[(B_{(1)} \otimes B_{(2)}) \otimes (B^\prime_{(1)} \otimes B^\prime_{(2)})\right]\\
&=& ( m\otimes m )\circ  \tau_{23}\circ(\Delta \otimes \Delta) \left(B\otimes B^\prime\right).\\
\end{eqnarray*}
\end{itemize}\endsquare\\\\
With the  help of  meromorphic extensions
of integrals of holomorphic radial symbols with linear constraints built in
the previous section, we build  a morphism from ${\cal H}$ into the algebra of
meromorphic functions. The following lemma follows from Corollary \ref{cor:meromultintsymb}.

. 
\begin{lem} \label{prop:meromultintsigma} Let ${\cal R}: \sigma\mapsto \sigma(z)$ be a holomorphic
  regularisation procedure on $CS(\R_+)$ and let $\sigma$  be a {\rm radial}
  classical  symbol    of order $a$ which is
sent via ${\cal R}$ to  $\sigma(z)$ of non constant affine order
$\alpha(z)$. The  map
$$\Phi_L^{{\cal R},\sigma}: B\mapsto 
\cutoffint_{\R^{nL}}\prod_{i=1}^I\sigma(z)\circ B(\xi_1, \cdots,\xi_L)\,
d\xi_1\cdots d\xi_L$$ yields a morphism of algebras
\begin{eqnarray*}
\Phi^{{\cal R},\sigma}: {\cal H}_L&\to&{\rm Mer}(\C)\\
 B&\mapsto& \Phi_L^{{\cal R}, \sigma}(B)
\end{eqnarray*}
i.e. 
$$\Phi^{{\cal R}, \sigma}(B\oplus B^ \prime)= \Phi^{{\cal R}, \sigma}(B)\, \Phi^{{\cal R}, \sigma}(B^\prime)\quad
\forall (B, B^\prime) \in {\cal H}^2.$$
\end{lem}
The following theorem then follows from Birkhoff factorisation combined with a
minimal substraction scheme
 along the lines of a general procedure described in \cite{M} (Theorem II.5.1).
\begin{thm}\label{thm:Birkhoffrenormalised} Let ${\cal R}$ be a continuous holomorphic  regularisation on
     $CS(\R^n)$ which sends a symbol of order $ a$   to a symbol of order
     $\alpha(z)= -q\, z+a$ for some $q>0$ and let  $\sigma\in CS(\R^n)$ be a
     radial symbol.  The map $\phi^{{\cal R}, \sigma}:= \Phi_+^{{\cal R},
       \sigma}(0)$  is  a character
\begin{eqnarray*}
\phi^{{\cal R}, \sigma}: {\cal H}&\mapsto & \C\\
              B  &\mapsto & \cutoffint_{\R^{nL}}^{{\cal R}, {\rm Birk}}
                \sigma^{\otimes I}\circ B
\end{eqnarray*}where $\Phi^{{\cal R}, \sigma}= \left(\Phi^{{\cal R}, \sigma}_-\right)^{*-1}
 \star \Phi^{{\cal R}, \sigma}_+$ is the unique Birkhoff decomposition of
 $\Phi^{{\cal R}, \sigma}$, $\star$ being the convolution product on the Hopf algebra.\\
 When $\sigma$ has order with real part $<-n$, the renormalised integral $ \cutoffint_{\R^{nL}}^{{\cal R}, {\rm Birk}}
                \sigma^{\otimes I}\circ B$ coincides with the ordinary
                integral  $ \int_{\R^{nL}}
                \sigma^{\otimes I}\circ B$.
\end{thm}
{\bf Proof:} By the very construction of the birkhoff factorised morphism, the multiplicativity  of
$\phi^{{\cal R}, {\rm Birk}}$ follows from the multiplicative property
of the morphism 
$\Phi^{{\cal R}, \sigma}$. The fact that the resulting renormalised integral  coincides with the ordinary integral 
$\int_{\R^{nL}}\tilde\sigma \circ B$ when $\sigma$ has order $a$
with real part $<-n$ follows from the fact that the map $\Phi^{{\cal R}, \sigma}$ is
then holomorphic around $0$ so that $\Phi^{{\cal R}, \sigma}_+= \Phi^{{\cal R}, \sigma}$. \endsquare
\subsection{Properties of renormalised multiple integrals with constraints}
By construction,  both {\rm renormalised} multiple integrals of
symbols with
constraints given by some matrix $B\in{\cal M}_{I, L}$, namely $ \cutoffint_{\R^{nL}}^{{\cal R},
{\cal E}}
              ( \sigma_1\otimes\cdots\otimes \sigma_I)\circ B$ obtained using
              evaluators, resp. 
$ \cutoffint_{\R^{nL}}^{{\cal R},
{\rm Birk}}
              \sigma^{\otimes I}\circ B$ obtained using Birkhoff factorisation
\begin{itemize}
\item factorise over disjoint sets of constraints:
\begin{eqnarray}\label{eq:renormalisedfactorisation}
\cutoffint_{\R^{n(L+L^\prime)}}^{{\cal R},
{\cal E}}
            ( \tilde  \sigma\otimes\tilde\sigma^\prime) \circ (B\oplus
            B^\prime)&=&\left(\cutoffint_{\R^{nL}}^{{\cal R},
{\cal E}}
             \tilde  \sigma \circ B\right)\, \cdot \,\left( \cutoffint_{\R^{nL^\prime}}^{{\cal R},
{\cal E}}
             \tilde  \sigma \circ
            B^\prime\right)\\
{\rm resp.}\quad \cutoffint_{\R^{n(L+L^\prime)}}^{{\cal R},
{\rm Birk}}
            ( \sigma^{\otimes I}\otimes \left(\sigma^\prime\right)^{\otimes I^\prime}) \circ (B\oplus
            B^\prime)&=&\left(\cutoffint_{\R^{nL}}^{{\cal R},
{\rm Birk }}
              \sigma^{\otimes I} \circ B\right)\, \cdot \,\left( \cutoffint_{\R^{nL^\prime}}^{{\cal R},
{\rm Birk }}
           \left(\sigma^\prime\right)^{\otimes I^\prime} \circ
            B^\prime\right). \nonumber
\end{eqnarray}
Here, $B\in {\cal M}_{I, L}(\R)$ and   $B^\prime\in {\cal M}_{I^\prime, L^\prime}(\R)$.
\item coincide with the corresponding ordinary integrals with constraints when
  the integrands lie in $L^1$:
 \begin{eqnarray}\label{eq:renormalisedcoincidence} \sigma_i\in L^1(\R
   n)\quad \forall i\in \{1, \cdots, I\}&\Rightarrow & \cutoffint_{\R^{nL}}^{{\cal R},
{\cal E}}
             (\sigma_1\otimes \cdots\otimes\sigma_I) \circ B=\int_{\R^{nL}}
              ( \sigma_1\otimes\cdots\otimes \sigma_I)\circ B\nonumber\\
{\rm resp.}\quad\sigma\in L^1(\R^n)&\Rightarrow&  \cutoffint_{\R^{nL}}^{{\cal R},
{\rm Birk }}
              \sigma^{\otimes I} \circ B= \int_{\R^{nL}}    \sigma^{\otimes I}\circ B
\end{eqnarray}
\end{itemize}
The following theorem shows that they moreover fulfill a covariance property
and hence obey a Fubini property.
\begin{thm}\label{thm:renormalisedcovariance} Let ${\cal R}: \sigma\mapsto \sigma(z)$ be a holomorphic
  regularisation procedure on $CS(\R_+)$ which sends a symbol $\tau$ of order
  $a$ to   ${\cal R}(\tau)(z)$ of non constant affine order
$-qz_i+a$, for some positive real number $q$ and let ${\cal E}$ be a
generalised 
evaluator at $0$  on  the algebra $ {\cal B}$ of meromorphic maps.\\
               For any $B\in {\cal M}_{I, L}(\R)$ of rank $L$, for any matrix
               $C\in GL_L(\R)$ and any radial classical symbols $\sigma_1,
               \cdots, \sigma_I, \sigma$ on $\R^n$ we have
\begin{eqnarray}\label{eq:renormalisedcovariance}
\cutoffint_{\R^{nL}}^{{\cal R}, {\cal E}}
              \left( (  \sigma_1\otimes \cdots\otimes\sigma_I)\circ B \right)\circ  C &=&\vert {\rm det} C\vert^{-1}\,  \cutoffint_{\R^{nL}}^{{\cal R}, {\cal E}}
               (  \sigma_1\otimes \cdots\otimes\sigma_I)\circ B\nonumber\\
{\rm resp.}\quad \cutoffint_{\R^{nL}}^{{\cal R}, {\rm Birk}}
              \left( \sigma^{\otimes I}\circ B \right)\circ  C &=&\vert {\rm det}
               C\vert^{-1}\,  \cutoffint_{\R^{nL}}^{{\cal R}, {\rm Birk}}
             \sigma^{\otimes I}\circ B.  
 \end{eqnarray}
 As a result, they obey a Fubini
                type property:
\begin{eqnarray}\label{eq:renormalisedFubini}
\cutoffint_{\R^{nL}}^{{\cal R}, {\cal E}}
                 (  \sigma_1\otimes \cdots\otimes\sigma_I)\circ
                 B(\xi_{\rho(1)}, \cdots, \xi_{\rho(L)})&=&\cutoffint_{\R^{nL}}^{{\cal R}, {\cal E}}
                 (  \sigma_1\otimes \cdots\otimes\sigma_I)\circ B(\xi_{1},
 \cdots, \xi_{L})\\
{\rm resp.}\quad \cutoffint_{\R^{nL}}^{{\cal R}, {\rm Birk}}
                 (  \sigma_1\otimes \cdots\otimes\sigma_I)\circ
                 B(\xi_{\rho(1)}, \cdots, \xi_{\rho(L)})&=&\cutoffint_{\R^{nL}}^{{\cal R}, {\rm Birk}}
                 (  \sigma_1\otimes \cdots\otimes\sigma_I)\circ B(\xi_{1},
 \cdots, \xi_{L})\quad \forall \rho\in \Sigma_L\nonumber.
\end{eqnarray}
\end{thm}
{\bf Proof:} The Fubini property follows from the covariance property choosing
$C$ to be a permutation matrix.
\\ Covariance  follows by analytic
continuation 
from the usual covariance property of the ordinary integral; indeed this leads
to the following  equalities of meromorphic
maps 
\begin{eqnarray*} \vert {\rm det} C\vert\, \cutoffint_{\R^{nL}}            \left( \left({\cal R}( \sigma_1)(z_1)\otimes \cdots \otimes {\cal R}(\sigma_I)(z_I)\circ B\right)\right)\circ  C &=&  \cutoffint_{\R^{nL}}\left( {\cal R}(\sigma_1)(z_1)\otimes \cdots \otimes {\cal R}(\sigma_I)(z_I)\right)\circ B\\
{\rm resp.}\quad \vert {\rm det} C\vert\, \cutoffint_{\R^{nL}}
           \left(  \left({\cal R}(\sigma)(z)\right)^{\otimes I}\circ B\right)\circ C &=&  \cutoffint_{\R^{nL}}
                 \left({\cal R}(\sigma)(z)\right)^{\otimes I}\circ B.
\end{eqnarray*} 
Applying a generalised evaluator ${\cal E}$ to either side of the first
equality or implementing Birkhoff factorisation to the morphisms arising on either side of the second
equality leads to the two identities of (\ref{eq:renormalisedcovariance}).  \endsquare
\vfill \eject \noindent
\section*{Appendix : Proof of Proposition \ref{prop:meroextmultint} }
To simplify notations, we set
$q_i(\underline \xi):= \sum_{l=1}^L b_{il}\, \xi_l$ where $\underline \xi:=
(\xi_1, \cdots, \xi_L)$ and  $b_i=-a_i$. For
Re$(b_i)$ chosen  sufficiently large,  we write
\begin{eqnarray*}
&{}&\int_{\left(\R^n\right)^L}
 \prod_{i=1}^I 
\langle q_i(\underline \xi)\rangle^{a_i}\, d\xi_1\cdots d\xi_L\\
&=&\frac{1}{\Gamma(b_1/2)\cdots \Gamma(b_I/2)} \int_0^\infty \e^{\frac{b_1}{2}-1}\cdots
 \e^{\frac{b_I}{2}-1}
\int_{\left(\R^n\right)^L}\, e^{-\sum_{i=1}^I \e_i\, 
\langle q_i(\underline \xi)\rangle^2} \,d\xi_1\cdots d\xi_L 
\end{eqnarray*} and 
 $$\sum_{i=1}^I \e_i\langle q_i(\underline \xi)\rangle^2=\sum_{l,
   m=1}^L\sum_{i=1}^I \e_i \, b_{i,l}b_{im} \xi_l\cdot
\xi_m+\sum_{i=1}^I \e_i=\sum_{l, m=1}^L\theta(\underline\e)_{lm} \xi_l\cdot \xi_m+\sum_{i=1}^I \e_i,$$
where $\xi_l\cdot \xi_m$ stands for the inner product in $\R^n$ and where we have set $$\theta(\underline\e)_{lm}:=\sum_{i=1}^I 
\e_i\, b_{il}b_{im}. $$ 
  Since the $\e_i$ are positive  $\theta(\e)$ is a non negative matrix,
  i.e. $ \theta(\e)(\underline \xi)\cdot \underline \xi\geq 0$. It is actually positive
  definite since
\begin{eqnarray*}
&{}&\sum_{l, m=1}^L\theta(\underline\e)_{l,m}
  \xi_l\cdot \xi_m=0\\
&\Rightarrow&\sum_{i=1}^I \e_i\vert q_i(\underline \xi)\vert^2=0
\Rightarrow q_i(\underline \xi)=0\quad \forall i\in \{1,
\cdots, I\}\\
&\Rightarrow&\sum_{i=1}^I \vert q_i(\underline \xi)\vert^2=\vert
B\underline  \xi\vert^2=0
\Rightarrow \underline \xi=0,
\end{eqnarray*}
using the fact that $B^*B$ is positive definite. The map
$\xi\mapsto \sum_{l, m=1}^L\theta(\underline\e)_{lm}\,  \xi_l\cdot \xi_m$ therefore defines a
positive definite quadratic form of rank  $L$.\\
A  Gaussian integration yields
$\int_{\left(\R^n\right)^L}
e^{-\sum_{i=1}^I\e_i\vert q_i(\underline \xi)\vert^2}\, d\xi_1\cdots d\xi_L= \left({\rm det}(\theta(\underline \e))\right)^{-n/2}.$
We want  to perform the integration over $\underline \e$: 
$$\frac{1}{\Gamma(b_1/2)\cdots \Gamma(b_I/2)}\int_0^\infty d \e_1\cdots 
\int_0^\infty d\e_I 
\e_1^{\frac{b_1}{2}-1}\cdots \e_I^{\frac{b_I}{2}-1}\left({\rm
    det}(\theta(\underline \e))\right)^{-\frac{n}{2}}\, e^{-\sum_{i=1}^n
  \e_i}.$$
Let us decompose the space $\R_+^k$ of parameters
$(\e_1, \cdots, \e_I)$ in regions $D_\tau$ defined by $\e_{\tau(1)}\leq \cdots \leq \e_{\tau(I)} $
for permutations $\tau\in \Sigma_I$. This splits the integral $\int_0^\infty d \e_1\cdots 
\int_0^\infty d\e_I 
\e_1^{\frac{b_1}{2}-1}\cdots \e_I^{\frac{b_I}{2}-1}\left({\rm
    det}(\theta(\underline \e))\right)^{-\frac{n}{2}}\, e^{-\sum_{i=1}^n
  \e_i} $  into
  a sum of integrals  $\int_{D_\tau} d \e_1\cdots 
 d\e_I 
\e_1^{\frac{a_1}{2}-1}\cdots \e_I^{\frac{a_I}{2}-1}\left({\rm
    det}(\theta(\underline \e))\right)^{-\frac{n}{2}}\, e^{-\sum_{i=1}^n
  \e_i}$. \\
 Let us focus on the integral over the domain $D$ given by $\e_1\leq \cdots \leq \e_k$; the
 results can then be transposed to other domains applying  a permutation\footnote{
Note that  a permutation $\tau\in \Sigma_I$ on the
$a_i$'s (and hence the $b_i$'s)  boils down to a permutation on the lines of the matrix
$(a_{il})$. Indeed, for any $\tau\in \Sigma_I$ 
\begin{eqnarray*}
&&\frac{1}{\Gamma(b_{\tau(1)})\cdots \Gamma(b_{\tau(I)})} \int_0^\infty d\e_1
\, \e_1^{\frac{b_{\tau(1)}}{2}-1}\cdots \int_0^\infty d\e_I\,
\e_I^{\frac{b_{\tau(I)}}{2}-1}\int_{\left(\R^n\right)^L} e^{-\sum_{i=1}^k
  \e_i\,\langle q_i(\underline \xi)\rangle^2}\\
&=&  \frac{1}{\Gamma(b_{1})\cdots \Gamma(b_{I})} \int_0^\infty d\e_1\cdots \int_0^\infty d\e_k\,
 \e_{\tau^{-1}(1)}^{\frac{b_1}{2}-1}\cdots
\e_{\tau^{-1}(k)}^{\frac{b_I}{2}-1}\int_{\left(\R^n\right)^L} e^{-\sum_{i=1}^k
  \e_{\tau^{-1}(i)}\,\langle q_{\tau^{-1}(i)}(\underline \xi)\rangle^2 }\\
&=&  \frac{1}{\Gamma(b_{1})\cdots \Gamma(b_{I})} \int_0^\infty d\e_1\cdots \int_0^\infty d\e_I\,
 \e_{1}^{\frac{b_1}{2}-1}\cdots
\e_{k}^{\frac{b_I}{2}-1}\int_{\left(\R^n\right)^L} e^{-\sum_{i=1}^k  \e_{i}\,\langle q_{\tau^{-1}(i)}(\underline \xi)\rangle^2 }.
\end{eqnarray*}
so that the $q_i$'s which determine the lines of the matrix are permuted. } 
 $b_i\rightarrow a_{\tau(i)}$ on 
 the $b_i$'s. 
We write the domain of integration as a union of cones
$0\leq \e_{j_1}\leq \cdots \leq \e_{j_I}$. For simplicity, we consider the region
$0\leq \e_1\leq \cdots \leq \e_I$ on which we introduce new variables $t_1,
\cdots, t_I$ setting 
$\e_i= t_{I}t_{I-1}\cdots t_i$. These new variables   vary in the domain
$\Delta:= \prod_{i=1}^{I-1} [0, 1]\times [0, \infty).$
Let us assume that $b_{il}=0$ for $i>i_l$, then the $l$-th line of $\theta$
reads
$$\theta(\underline\e)_{lm}=\sum_{i=1}^I 
t_I\cdots t_i \, b_{il}b_{im}=\sum_{i=1}^{i_l} 
t_{I}\cdots t_{i} \, b_{il}b_{im}= t_{I}\cdots t_{i_l}\left(b_{i_ll}b_{i_lm}+\sum_{i=1}^{i_l-1} 
t_{i_l-1}\cdots t_{i} \, b_{il}b_{im}\right) $$
or equivalently the $m$-the column of $\theta$ reads
$$\theta(\underline\e)_{lm}=\sum_{i=1}^I 
t_I\cdots t_i \, b_{il}b_{im}=\sum_{i=1}^{i_m} 
t_{I}\cdots t_{i} \, b_{il}b_{im}= t_{I}\cdots t_{i_m}\left(b_{i_ll}b_{i_lm}+\sum_{i=1}^{i_m-1} 
t_{i_m-1}\cdots t_{i} \, b_{il}b_{im}\right).$$
Factorising out $\sqrt{t_{I}\cdots t_{i_l}}$ from the $l$-th  row and 
$\sqrt{t_{I}\cdots t_{i_m}}$ from the $m$-th column for every $l,m\in[[1, L]]$
produces a symmetric matrix $\tilde \theta(\underline t)$. \\
Following \cite{Sp} we show that its determinant
does not vanish on the domain of integration; if it did vanish at some point
$\underline \tau$, $\tilde \theta(\underline \tau)$ would define a non
injective map $\tilde \theta(\underline
\tau): (x_1, \cdots, x_L)\mapsto \left(\sum_{l=1}^L\tilde \theta(\underline
\tau)_{1l}x_l, \cdots,  \sum_{l=1}^L\tilde \theta(\underline
\tau)_{Ll}x_l \right)$, i.e. there would be some non zero  $L$-tuple
$\underline w:=(x_1, \cdots,
x_L)\in\R^L$ such that $\tilde \theta(\underline
\tau)(\underline x)=0$ which would in turn imply that 
 $ \sum_{l=1}^L\sum_{m=1}^L  x_l\left(\tilde \theta(\underline
\tau)\right)_{lm}x_m =\underline  x\cdot \tilde \theta(\underline
\tau)(\underline x)=0$. From there we would infer that 
\begin{eqnarray}\label{eq:S1}&{}&\sum_{l=1}^L \sum_{m=1}^M \sum_{i=1}^I \tau_I\cdots \tau_i \, b_{il} b_{im}
x_l x_m= \sum_{i=1}^I \left( \sum_{l=1}^L \sqrt{ \tau_I\cdots
    \tau_i}\,b_{il}x_l\right)^2=0\nonumber\\
&\Longrightarrow &\sum_{l=1}^L
\sqrt{\tau_I\cdots\tau_i}\, b_{il} x_l=0\\
&\Longrightarrow &\sum_{l=1}^L\left(b_{i_ll} x_l +
\sqrt{\tau_{i_{l-1}}\cdots\tau_i}  \, b_{il} x_l\right)=0\quad \forall i\in [[1, I]],\end{eqnarray}
where we have factorised out $\tau_I\cdots\tau_{i_l}$ in the last expression.
Let us as in \cite{Sp} choose  $M={\rm max}\{l, x_l\neq 0\}$; in particular
$l>M\Rightarrow x_l=0$. On the other hand, since
$l<M\Rightarrow i_l<i_M$ we have $l<M\Rightarrow b_{i_Ml}=0$. Choosing
$i=i_M$ in (\ref{eq:S1}) reduces the sum to one term $b_{i_MM} x_M$ which
would therefore vanish, leading to a contradiction since neither $b_{i_MM}$
nor $x_M$ vanish by assumption.\\
We thereby conclude that ${\rm det}\tilde \theta(\underline t)$ does not vanish on
the domain of integration.
\\
Performing the change of variable $(\e_1, \cdots, \e_I)\mapsto (t_1, \cdots, t_I)$
in the  integral, which introduces a jacobian determinant $\prod_{i=1}^I
t_i^{i-1} $, we write   the integral:
\begin{eqnarray}\label{eq:S2}
&{}&\frac{1}{\Gamma(b_1/2)\cdots \Gamma(b_I/2)}\int_0^1 d t_1\cdots 
\int_0^1 dt_{I-1}\int_0^\infty dt_I\prod_{i=1}^I
t_i^{i-1} \prod_{l=1}^L(t_I\cdots t_{i_l})^{-\frac{n}{2}}\nonumber\\
 &{}& \cdot
 \prod_{i=1}^{I}(t_I\cdots t_{i})^{\frac{b_i}{2}-1}\, e^{-\sum_{i=1}^I
   t_I\cdots t_i}\, \left({\rm det}\tilde \theta (t)\right)^{-n/2}\nonumber \\
 &=&\frac{1}{\Gamma(b_1/2)\cdots \Gamma(b_I/2)}\int_0^\infty dt_I
  \int_0^1 d t_1\cdots 
\int_0^1 dt_{I-1}\, \nonumber \\
&& \prod_{i=1}^I
t_i^{\frac{b_1+\cdots +b_i}{2}-1}  (t_I\cdots t_{i_L})^{-n\frac{L}{2}} 
(t_{{i_L}-1}\cdots t_{i_{L-1}})^{-n\frac{L-1}{2}} \cdots (t_{i_2}\cdots
t_{i_1})^{-\frac{n}{2}}\, h(\underline t)\nonumber\\
&=&\frac{1}{\Gamma(b_1/2)\cdots \Gamma(b_I/2)}\int_\Delta dt_I\cdots
   d t_1\,\prod_{i=1}^I
t_i^{\frac{b_1+\cdots +b_i-ns_i}{2}-1} \, h(\underline t)
 \end{eqnarray}
where the $s_i's$ are positive integers depending on the size and shape of the
matrix 
$B$ (via the  $i_l$'s) \footnote{The integers $s_i$'s do not  depend on the explicit coefficients of the
matrix.We have
  $i_l\geq l$ so that $s_i\leq i$; in particular,   ${\rm
    Re}(a_i)<-n\Rightarrow{\rm Re}(b_1)+\cdots +{\rm Re}(b_i)-ns_i\geq {\rm
    Re}(b_1)+\cdots +{\rm Re}(b_i)-n\, i>0$ so that as expected, the above integral converges. } and where  we have set $$h(\underline t):=  e^{-\sum_{i=1}^I t_I\cdots
  t_i}\,\left( {\rm det}\tilde \theta (t)\right)^{-n/2}=\left({\rm det}\tilde \theta (t)\right)^{-n/2}\,
\prod_{i=1}^I  e^{- t_I\cdots t_i}.$$
 Since ${\rm det}\tilde \theta (t)$ is polynomial in the $t_i$'s, the convergence of the integral in $t_I$ at
infinity is taken care of by the function $e^{- t_I\cdots t_1}$ arising in $h$.
On the other hand, $h$ is smooth on the domain of integration since it is
clearly smooth outside the set of points for which ${\rm det}
\tilde \theta (\underline t)$ vanishes, which we saw is a void set. Thus,     the various integrals converge at $t_i=0$  for
Re$(b_i)$ sufficiently large. \\ Integrating by
parts with respect to each  $t_1, \cdots, t_{I}$ introduces factors
$\frac{1}{b_1+\cdots+ b_i-n\,s_i+2m_i}, m_i\in \N_0$ when taking primitives of $
t_i^{\frac{b_1+\cdots +b_i-n\,s_{i}}{2}-1}$ and differentiating $h(\underline
t)$.\\
We thereby build a meromorphic extension $\cutoffint_{\R^{nL}}
 \prod_{i=1}^k\langle  \sum_{l=1}^L b_{il} \xi_l\rangle^{a_i}$ to the whole complex
plane as a sum over permutations $\tau\in \Sigma_I$ of expressions:
$$\frac{1}{\prod_{i=1}^I \Gamma(b_i)}\,\left(
\frac{\int_{\Delta} \prod_{i=1}^kt_i^{\frac{b_{\tau(1)}+\cdots +b_{\tau(i)}-n\, s_{i\tau(i)}}{2}+m_i} \, h_\tau^{(m_1+\cdots+ m_I)}(\underline t)}{\prod_{i=1}^I\left((b_{\tau(1)}+\cdots+b_{\tau(i)}-n\,s_{\tau, i} )\cdots (b_{\tau(1)}+\cdots+
  b_{\tau(i)}-n\,s_{\tau, i}+2m_i )\right)}+{\rm boundary}\quad {\rm terms}\right)\,
 $$ where the  boundary terms
 on the domain $\Delta$ are produced by the iterated $m_i$ integrations by
 parts in each variable $t_i$. Here $s_{\tau,i}\leq i$ is a positive integer
 depending on $\tau$ and the shape of the matrix and 
 we have chosen the $m_i$'s sufficiently large for the term
 $\int_{\Delta} \prod_{i=1}^I
t_i^{\frac{b_{\tau(1)}+\cdots +b_{\tau(i)}-ns_{\tau,i}}{2}+m_i} \, h_\tau^{(m_1+\cdots+ m_I)}(\underline t)$  to
converge. The boundary terms are of the same type, namely they are
proportional to 
 $$\frac{\int_{\Delta^\prime} \prod_{i=1}^I
t_i^{\frac{b_{\tau(1)}+\cdots +b_{\tau(i)}-ns_{\tau,i}}{2}+m_i^\prime} \, h^{(m_1^\prime+\cdots+ m_k^\prime)}(\underline t)}{\prod_{i=1}^k\left((b_{\tau(1)}+\cdots+b_{\tau(i)}-n\,s_{\tau, i}  )\cdots (b_{\tau(1)}+\cdots+  b_{\tau(i)}-n\,s_{\tau, i} +2m_i^\prime )\right)}$$ for some domain
  $\Delta^\prime=\prod_{i=1}^{I^\prime-1}[0,1 ]\times [0, \infty[$  for some $I^\prime < I$ or
  $\Delta^\prime=\prod_{i=1}^{I^\prime-1} [0, 1]$ for some $I^\prime \leq I$  and some non
  negative integers $m_i^\prime\leq m_i$ with at least one
  $m^\prime_{i_0}<m_{i_0}$. \\\\
 This produces a meromorphic map which on the domain $ \cap_{i=1}^I\{{\rm Re}(b_{\tau(1)}+\cdots+
  b_{\tau(i)}) +2m_i>ns_{\tau, i}\}$ reads $$\frac{1}{\prod_{i=1}^I \Gamma(b_i)}\,\sum_{\tau\in \Sigma_I}\, \frac{H_{\tau, \underline m}(b_{1}, \cdots,
b_{I})}{\prod_{i=1}^I\left((b_{\tau(1)}+\cdots+
  b_{\tau(i)}-n\,s_{\tau,i}  )\cdots (b_{\tau(1)}+\cdots+
  b_{\tau(i)}-n\,s_{\tau,i} +2m_i) \right)}$$
with $H_{\tau, \underline m}$ holomorphic on that domain. It therefore extends
to a meromorphic map on the whole complex space with  simple 
simple  poles on   a countable set of affine
hyperplanes $\{a_{\tau(1)}+\cdots
+a_{\tau(i)}+ns_{\tau,i}\in 2\N_0 \}$, where as beofre,  the $s_{\tau,i}$'s are integers which depend 
on the permutation $\tau$ and on the size  $L\times I$  shape (i.e. on the $l_i$'s) but not on the actual
coefficients of the matrix.
\\  Let us further observe that since $s_{\tau, i}\leq i$, if ${\rm Re}(a_i)<-n\Rightarrow {\rm
    Re}(b_i)>n$ for any $i\in\{1, \cdots, I\}$, then for any $\tau\in \Sigma_I$
we have ${\rm Re}(b_{\tau(1)}+\cdots+
  b_{\tau(i)})-n\,s_{\tau,i}>0$ so that  we recover the fact that the map $(a_1,
  \cdots, a_I)\mapsto  \cutoffint_{\R^{nL}}
 \prod_{i=1}^k\langle  \sum_{l=1}^L b_{il} \xi_l\rangle^{a_i}$ is holomorphic
 on the domain $D:=\{\underline
 a=(a_1, \cdots, a_I)\in \C^I,  {\rm Re}(a_i)<-n, \quad \forall
 i\in\{1, \cdots, I\}\}$. 
\endsquare
\vfill \eject 

\vskip 10mm {\small \noi\textsc{Laboratoire de Math\'ematiques,
Complexe des C\'ezeaux, Universit\'e Blaise Pascal, 63 177
Aubi\`ere Cedex F. E-mail:} 
sylvie.paycha@math.univ-bpclermont.fr}

\noindent
\begin{thebibliography}{99}
\bibitem[BM]{BM} L. Boutet de Monvel, {\it Alg\`ebres de Hopf des diagrammes de
    Feynman, renormalisation et factorisation de Wiener-Hopf} (d'apr\`es
  Connes et Kreimer), S\'eminaire Bourbaki, Ast\'erisque {\bf 290} (2003)
  149-165
\bibitem[BP]{BP} N. Bogoliubov, O. Parasiuk, {\" Uber die Multiplikation der
    Kausalfunktionen in der Quantentheorie des Felder},  Acta Math. {\bf 97}
  (1957) 227-265
\bibitem[BW1]{BW1} Ch. Bogner, S. Weinzierl, {\it  Resolution of singularities
  for multi-loop integrals}, hep-th 0709.4092v2

\bibitem[BW2]{BW2} Ch. Bogner, S. Weinzierl, {\it Periods and Feynman
    integrals}, hep-th 0711.4863.v1
\bibitem[CaM]{CaM} C. de Calan, A. Malbouisson, {\it Infrared and untraviolet
    dimensional meromorphy of Feynman amplitudes}, Comm. Math. Phys. {\bf 90}
  413-416 (1983)
\bibitem[CK]{CK} A. Connes, D. Kreimer, \otherterm{ Hopf algebras,
    Renormalisation and Noncommutative Geometry}, Comm. Math. Phys.{\bf  199} (1988)
  203-242 
\bibitem[CM]{CM} A. Connes, M. Marcolli, {\it Noncommutative Geometry, Quantum
  Fields  and Motives}, to appear
\bibitem[C]{C} J. Collins, {\it Renormalization: general theory};
  arXiv:hep-th/0602121v1 (2006) and {\bf Renormalization}, Cambridge
  Univ. Press (1984)
\bibitem[E]{E} P. Etingof, {\it  A note on dimensional regularization} in 
 {\it Quantum Fields and Strings: A Course for Mathematicians}, Amer.
 Math. Soc.  
(2000) 597-607
\bibitem[E-FGK]{E-FGK} K. Ebrahimi-Fard, Li Guo, Dirk Kreimer, {\it Spitzer's
    identity and the algebraic Birkhoff decomposition in QFT},
  arXiv:hep-th/0407082 (2004)
  \bibitem[G]{G} V. Guillemin, {\it   A new proof of
Weyl's formula on the asymptotic distribution of eigenvalues}, Adv.
Math. {\bf 55} (1985) 131--160
\bibitem[H]{H} K. Hepp, {\it  Proof of the Bogoliubov-Parasiuk Theorem on
  renormalization}, Comm. Math. Phys. {\bf 2} (1966) 301-326
\bibitem[HV]{HV} G. t'Hooft, M. Veltman, {\it Regularisation and
    renormalisation of gauge fiels, Nuclear Physics {\bf B44} (1972) 189-213}
  \bibitem[K]{K} Ch. Kassel,\otherterm{ Le r\'esidu non commutatif [d'apr\`es
    Wodzicki]}, S\'em. Bourbaki {\bf 708} (1989) 
    \bibitem[KV]{KV} M. Kontsevich, S. Vishik,\otherterm{Geometry of determinants of elliptic operators}, Func.
Anal. on the Eve of the XXI century, Vol I, Progress in
Mathematics {\bf 131} (1994)  173--197 ; \otherterm{ Determinants
of elliptic pseudodifferential operators}, Max Planck Preprint
(1994)
\bibitem[Kr ]{Kr} D.Kreimer, D. Kreimer, \otherterm{ On the Hopf algebra of perturbative 
quantum field theory}, Adv. Theo. Math. Phys.{\bf  2} (1998) 303-334 
\bibitem[L]{L} M. Lesch, \otherterm{
On the non commutative residue for pseudo-differential operators
with log-polyhomogeneous symbols}, Ann.  Global  Anal.
Geom. {\bf 17}(1998)  151--187
\bibitem[LP]{LP} M. Lesch, M. Pflaum, {\it Traces on algebras of parameter
    dependent pseudodifferential operators and the eta-invariant},
  Trans. Amer. Soc. {\bf 352} n.11 (2000) 4911-4936 
\bibitem[M]{M} D. Manchon, {\it Hopf algebras, from basics to applications to
    renormalization}, Glanon Lecture Notes,  2002
\bibitem[MP]{MP} D. Manchon, S. Paycha, {\it Shuffle relations for regularised
integrals of symbols}, Commun. Math. Phys. {\bf 270} 13-51 (2007)
\bibitem[MMP]{MMP} Y. Maeda, D.Manchon, S. Paycha, \otherterm{ Stokes'
    formulae on classical symbol valued forms and applications}, Preprint
  2005 
\bibitem[P1]{P1} S.Paycha, {\bf Regularised sums, integrals and traces; a
    pseudodifferential point of view},
  Lecture Notes in preparation
  (http://www.lma.univ-bpclermont.fr/~paycha/publications/html)
\bibitem[P2]{P2} S. Paycha, {\it The noncommutative residue and the canonical
    trace in the light of Stokes' and continuity properties} arXiv:0706.2552
  (2007)
\bibitem[P3]{P3}  S. Paycha, {\it Discrete sums of classical symbols on $\Z^d$ and zeta functions
  associated with Laplacians on tori} arXiv:0708.0531 (2007)
 \bibitem[P4]{P4}  S.Paycha, {\it Renormalised integrals and sums with constraints; a
     comparative study}, work in progress
\bibitem[PS]{PS} S. Paycha, S. Scott, {\it A  Laurent expansion for regularised integrals of
holomorphic symbols}, Geom.  Funct. Anal. {\bf 17} (2007) 491-536
\bibitem[Sm1]{Sm1} V.A. Smirnov, {\it Infrared and ultraviolet divergences of
    the coefficient functions of Feynman diagrams as tempered distributions I}
 Theoretical and Mathematicial Physics  1981 {\bf 44:3} 
 761-773 transl. from Teor. Mat. Fiz. {\bf 44} 307-320 (1980)
\bibitem[Sm2]{Sm2} V.A. Smirnov, {\bf 
   Evaluating Feynman integrals}, Springer Tracts in Modern Physics  211 (2004)\bibitem[Sm3]{Sm3} V.A. Smirnov,  {\bf
   Renormalization and asymptotic expansions}, Birkh\"auser 1991
\bibitem[Sp]{Sp} E. Speer, {\it Analytic renormalization},
  Journ. Math. Phys. {\bf Vol 9} 1404--1410 (1968)
\bibitem[W]{W} M. Wodzicki, { \it Non
commutative residue} in Lecture Notes in Math. {\bf 1283},
Springer Verlag 1987
\end{thebibliography}
\end{document}